\documentclass[usenatbib,referee]{mn2e}

\usepackage{epsf}
\usepackage{graphicx}
\usepackage{natbib}
\newcommand{\bmth}[1]{\mbox{\boldmath${#1}$}}

\title[Inertial waves in rotating bodies: a WKBJ formalism for  inertial modes ]
{Inertial waves in rotating bodies: a WKBJ formalism for  inertial modes and 
a comparison with numerical results} 
\author[P. B. Ivanov and
J. C. B. Papaloizou]{
P. B. Ivanov$^{1,2}$\thanks{E-mail:pbi20@cam.ac.uk
(PBI) J.C.B.Papaloizou@damtp.cam.ac.uk (JCBP)} 
and J. C. B. Papaloizou$^{1}$
\footnotemark[1]\\
$^{1}$Department of Applied Mathematics and Theoretical Physics,
University of Cambridge,\\
Centre for Mathematical Sciences, 
Wilberforce Road, Cambridge, CB3 0WA, UK \\
$^{2}$Astro Space Centre, P. N. Lebedev Physical Institute,
4/32 Profsoyuznaya Street,
Moscow, 117810, Russia}

\begin{document}

\date{Accepted Received ; in original form }

\pagerange{\pageref{firstpage}--\pageref{lastpage}} \pubyear{2002}

\maketitle

\label{firstpage}

\begin{abstract}
Inertial waves governed by Coriolis forces may play an important
role in several astrophysical settings, such as eg.  tidal
interactions, which  may occur in extrasolar planetary systems
and close binary systems, or in rotating compact objects emitting
gravitational waves. Additionally, they are of interest in other
research fields, eg. in geophysics.

However, their analysis is complicated by the fact that
in the inviscid case the normal mode spectrum is either everywhere dense
or continuous in any frequency interval contained within the inertial
range. Moreover, the equations governing the corresponding eigenproblem are, in
general, non-separable. 

In this paper we develop a consistent WKBJ formalism, together
with a formal first order perturbation  theory
for calculating the properties of the normal modes  of a uniformly rotating coreless
body (modelled as a polytrope and referred hereafter to as a planet) 
under the assumption of a spherically symmetric structure.
The eigenfrequencies, spatial form of the associated
eigenfunctions  and  other properties we  obtained analytically 
using the WKBJ eigenfunctions are in good agreement with corresponding
results obtained by numerical means  for a variety of planet
models even for global modes with a large scale distribution of
perturbed quantities. This indicates that even though they are embedded in a dense
spectrum,  such modes can be identified and followed
as model parameters changed and that first order perturbation
theory can be applied.

This is used to estimate corrections
to the eigenfrequencies as a consequence of the anelastic
approximation,  which we argue here
to be small when the rotation frequency is small. These are  compared
with simulation results in an accompanying paper with a good agreement  
between theoretical and numerical results.

The results reported here may provide a basis of theoretical
investigations of inertial waves in many astrophysical and other 
applications, where a rotating body can be modelled as a 
uniformly rotating barotropic object, for which the density has, close to its surface,
an approximately power law dependence on distance from the surface.

\end{abstract}

\vspace{-1cm}
\section{Introduction}\label{intro}

In astrophysical applications inertial waves that can exist in rotating bodies  
may be excited by several different physical mechanisms, most notably 
through tidal perturbation by a companion (eg. Papaloizou \& Pringle 1981,
hereafter PP) or in the case of compact objects through secular instability arising through
gravitational wave losses (eg. Chandrasekhar 1970, Friedman \& Schutz
1978, Andersson 1998, Friedman $\&$ Morsink 1998). They also can play
a role in other physical systems. For example, they can also be excited
by several mechanisms in the Earth's  fluid core with possible detection
being announced (Aldridge $\&$ Lumb 1987).

For  rotating planets and stars  that have  a barotropic
equation of state these  wave modes
are governed by Coriolis forces and so have oscillation periods that
are comparable to the rotation period. They are accordingly readily excited 
by tidal interaction with a perturbing body when the characteristic time associated
with the orbit is comparable to the rotation period, which is expected naturally
when the rotation period and orbit become tidally coupled.
They may then play an important role in governing the secular orbital evolution
of the system.

Inertial modes excited in close binary systems in circular orbit were considered by
PP and Savonije \& Papaloizou  (1997). Wu (2005)a,b considered the excitation of
inertial modes in Jupiter as a result of tidal interaction with a satellite
and excitation as a result of  a parabolic  encounter of a planet or star
 with a central star was studied by Papaloizou \& Ivanov (2005), hereafter referred to as PI
and Ivanov \& Papaloizou (2007), hereafter referred to as  IP.
The latter work was applied to the problem of circularisation
of extrasolar giant planets starting with high eccentricity.  
In that  work the planet  was  assumed coreless. Ogilvie \& Lin  (2004) and
Ogilvie (2009) have considered the case of a cored planet in circular orbit
around a central star and found that inertial waves play an important
role.

The importance of the role played by  inertial waves in the  transfer of the rotational energy of a rotating
neutron star to gravitational waves via the
Chandrasekhar-Friedman-Schutz  (CFS) instability was pointed out by Andersson
(1998). Later studies mainly concentrated on physical mechanisms of
dissipation of energy stored in these modes that limit amplitudes of
the modes, and, consequently, the strength of the  gravitational wave
signal. In these studies either numerical methods or simple local
estimates of properties of inertial modes were mainly used, see eg. 
Kokkotas (2008) for a recent review and references.

An analytical treatment of problems related to inertial waves, such as
eg. finding normal mode spectra and  eigenfunctions, and  
coupling them  to other physical fields, etc.,   
is difficult due to a number of  principal complicating technical
issues.

In particular, the dynamical equations governing the  perturbations of
a rotating body (called planet later on) are, in general, non-separable, for compressible
fluids. When such fluids are considered and rotation is assumed to be
small,  a low frequency anelastic approximation 
that filters out the high frequency modes is often used (see eg. PP).
This simplifies the problem to  finding solutions to leading 
order in the  small parameter $R_{*}^{3}\Omega^{2}/(GM)$, where $\Omega $ is
the rotation frequency, $G$ is the constant of gravity and $M_{*}$,
$R_{*}$ are the mass and radius of the planet. In this approximation
eigenfrequencies of inertial modes are proportional to $\Omega ,$
while the form  of the spatial distribution  of  perturbed
quantities does not depend on the rotation rate. However, even when
this approximation is adopted, the problem is, in general,
non-separable apart from models with a  special form of  density  distribution,
 see Arras et al (2003), Wu (2005)a and below.

Additionally, the problem of   calculating the inertial mode spectrum
and its response to tidal forcing is complicated by the fact that
in the inviscid case  the spectrum is either everywhere dense or continuous
in any frequency interval it spans (Papaloizou \& Pringle, 1982). This 
is in contrast  to the situation of, for example, high frequency $p$ modes,
which are discrete with well separated eigenvalues.
When the anelastic approximation  is adopted
the singular ill posed nature of the inviscid eigenvalue problem is seen
to come from the fact that the  governing equation is hyperbolic 
and the nature of the spectrum is  determined by the properties
of the characteristics (eg. Wood 1977).  A discrete spectrum is believed to occur
when there are no such trajectories that define  periodic attractors.
Otherwise the  inviscid spectrum is continuous. Then, when a small viscosity
is introduced the spectrum becomes  discrete but normal modes
have energy focused onto wave attractors (see eg. Ogilvie \& Lin 2004).
Given these complexities it is desirable to work with and compare
a variety of analytical and numerical approaches.  

Coreless  inviscid rotating planets with an assumed spherical or ellipsoidal
shape have a discrete but everywhere dense spectrum that makes difficulties
for example with mode identification and application of  standard perturbation theory.
However, numerical work indicates that there are well defined global modes
that can be identified and followed through a sequence of models  
(eg.  Lockitch \& Friedman, 1999, hereafter  LF, and   PI).
In this paper we investigate the inertial mode spectrum
of a uniformly rotating  coreless barotropic  planet or star 
and its tidal response by a WKBJ
approach coupled with first order perturbation
theory   and compare its eigenvalue predictions with numerical results obtained by a variety
of authors  and find good agreement  apart from some unidentified WKBJ modes that
are near the limits of the spectrum and for which the perturbation theory appears not to work.
For the identified modes we also find remarkably  good agreement for the form of the eigenfunctions. 
This indicates that they can be represented at  low resolution
with  small scale phenomena being unimportant,  meaningful 
mode identification (in that the modes can be followed
from one model to another) and at least first order perturbation theory works for these modes.
 
 This is also confirmed in a following paper  (hereafter referred to as PIN)
 where we investigate the inertial mode
 spectrum and its tidal response by numerical solution of an initial value problem
 {\it  without the anelastic approximation}.
 We are able to confirm the validity of the anelastic approximation
 and the applicability of the first order perturbation theory developed here  for demonstrating this
 as well as estimating eigenvalues.
 Thus a suggestion  of Goodman \& Lackner (2009) that tidal interaction
 might be seriously overestimated by use of the anelastic approximation
 is not confirmed.
 
A WKBJ approach to the same problem was also considered by  Arras
et al (2003) and Wu (2005)a. However, in this work  only terms
of leading order in an expansion in inverse powers of a large WKBJ
parameter $\lambda $ (see the text below for its definition) were taken
into account and treatment of perturbations near the surface  and
close to the rotational axis were oversimplified.
 As a consequence,
although their results  are correct in the 
formal limit $\lambda \rightarrow \infty, $ they cannot be used 
to make a correspondence between WKBJ modes  and those  obtained  numerically,
or an approximate description of modes with a scale that is not very small.
 In this paper we treat the
problem in a more  extended way,  considering  terms of the next
$O(\lambda^{-1})$ order together with an accurate treatment of
perturbations near the surface and close to the rotation
axis. Additionally, we consider a frequency correction of the next
 order, $O(\lambda^{-2}),$  for modes having non-zero azimuthal number,
$m$.

We checked results obtained with use of the WKBJ formalism against
practically all numerical data existing in the literature finding good
agreement in practically all cases. Therefore, we can assume that our 
formalism may be applied to provide an approximate analytic description of
inertial modes,  including those with large scale variations,
where the WKBJ approach might be expected to be  invalid. 
Also, different quantities associated with the modes may be described
within the framework of our formalism or its natural extension, such as 
the tidal overlap integrals (see PI and IP), quantities determining
the growth rate due to the CFS instability and decay of inertial waves due 
different processes, eg. by non-linear mode-mode interactions (see
eg. Schenk et al 2002, Arras et al 2003). Thus, the formalism developed
here may provide a basis for the  analytic  treatment of inertial waves in 
many different astrophysical applications.

The plan of the paper is as follows.
In section  \ref{sec2} we briefly review the basic equations and their linearised form
for a uniformly rotating barotropic planet or star. In section \ref{anelastic} 
we go on to consider these in the anelastic approximation which is appropriate when the
rotation frequency of the star is very much less than the critical
or break up rotation frequency. We give a simple physical argument
why we expect this approximation to be valid in this limit even when
the sound speed tends to a small value or possibly
zero at the surface of the configuration.
In section \ref{sec2.5} we 
give a brief discussion about when discrete normal modes may
be expected to occur such as in the case of a coreless
slowly rotating planet with surface boundary assumed to be
either spherical  or ellipsoidal.
We then  present a formal first order perturbation
theory that can be used to estimate corrections to eigenfrequencies
occurring as either a consequence of  terms neglected in the WKBJ approximation
or the anelastic approximation. The latter application is tested
by a direct comparison with the results of numerical simulations
in PIN. Section \ref{sec2.6} concludes with a brief account of
the form of the anelastic equations in pseudo-spheroidal coordinates 
in which they become separable for density profiles of the form
$\rho \propto (1-r^2/R_*^2)^{\beta},$ where $r$ is the local radius,
$R_*$ is the surface radius and $\beta$ is a constant.
 (Arras et al 2003, Wu 2005a).

In section \ref{sec3} we develop a WKBJ approximation for calculating 
the normal modes which is based on the idea that in the short wavelength
limit these modes coincide with those
appropriate to separable cases which include
the  homogeneous incompressible sphere as a well known example.
Solutions of a general WKBJ form appropriate to the interior
of the sphere are matched to solutions appropriate to the 
surface regions where they become separable which is the case
when the density vanishes as a power of the distance to the boundary
as is expected for a polytropic equation of state.
This matching results in an  expression for the eigenfrequencies
given in section \ref{sec3.5}.

In section \ref{surface} we go on to 
develop expressions for the eigenfunctions appropriate
to  any location in the planet including the rotation axis and the
critical latitude region where one of the inertial mode characteristics
is tangential to the planet surface. These solutions are then
used to obtain corrections to the eigenfrequencies resulting 
from density gradient terms neglected in the initial WKBJ  approximation
in section \ref{sec3.9}.
In section \ref{sec4} we compare the corrected  eigenfrequencies obtained from
the WKBJ approximation with those obtained numerically by several different
authors who used differing numerical approaches and find good agreement
even for global modes. A similar comparison with the results of  numerical
simulations for a polytropic model
with  positive results  is reported in PIN.
We also  compare the forms of the eigenfunctions 
with those obtained 
in Ivanov \& Papaloizou (2007) 
and find a  good agreement 
even for global modes.

{Finally in section \ref{sec5.1} we discuss our 
results in the context of the
evaluation of the overlap integrals that occur in evaluating the
response to tidal forcing. We show that 
these vanish smoothly in the limit  that the polytropic index tends to zero
and we indicate that they vanish at the lowest WKBJ order
and  are  thus expected to vanish rapidly as the order of the mode increases.
We go on to summarize our conclusions in section \ref{Conclu}.}

\section{Basic definitions and equations}\label{sec2}

In this section we review the formalism and equations we adopt in this paper.
As much of this  has been presented 
in  previous work  (PI, IP) only a brief review is given here.

In what follows we continue to investigate oscillations of  a uniformly rotating 
fully convective body referred hereafter to as a planet, 
focusing on the low frequency branch associated with inertial waves.

\subsection{Framework for linear perturbation analysis}

The planet is characterised by its mass $M_{*}$, radius $R_{*}$ and
the associated characteristic frequency 
\begin {equation}
\Omega_{*}=\sqrt{GM_{*}\over R_{*}^{3}}, \label{eqn p2} 
\end{equation}
where $G$ is the gravitational constant.  We adopt a cylindrical coordinate system
$(\varpi, \phi, z)$ and associated  spherical coordinate system
$(r, \phi, \theta)$ with origin at the centre of mass of the planet.

In this paper we make use of the
Fourier transform of a general perturbation 
quantity, say Q,  with respect to  the azimuthal angle  $\phi$ and the time $t$ in
the form
\begin{equation}
Q = \sum_{m}\left(  \exp({im\phi})\int^{+\infty}_{-\infty}d\sigma \tilde
Q_{m}\exp({-i\sigma t}) + cc \right ), \label{eq p1} \end{equation}
where the sum is over $m=0$  and $2$ and $cc$ denotes
the complex conjugate of the preceding quantity  hereafter. 
The reality of $Q$
implies that the   Fourier transform, indicated by tilde  satisfies 
$\tilde Q_{m}(\sigma) = \tilde
Q_{-m}^*(-\sigma).$
The inner products of two complex
scalars $Y_{1}$, $Y_{2}$, 
that are functions of $\varpi,$  $z$ and $t$  are defined as
\begin{equation}
(Y_{1}|Y_{2})=\int _S\varpi d\varpi dz
(Y_{1}^{*}Y_{2}), \label{eq p2} 
\end{equation}
where $*$ denotes 
the complex conjugate. Note that the definition of the inner product differs from 
what is given in IP where the planet's density $\rho $ was used as a weight function. 
Integrals  of this type are
always  taken over  the  section  $S$
of the unperturbed planet  for which $\phi =0.$

\subsection{Linearised equations of motion governing the response to tidal perturbation}

We assume that the planet is rotating with uniform angular velocity
$\bmth {\Omega }$. 
The hydrodynamic equations for the perturbed quantities 
take the simplest form in the rotating
frame with $z$ axis along the  direction of rotation.

Since the planet is fully convective,
the entropy per unit of mass of the planetary gas remains
approximately the same over the volume of the planet, 
and the pressure $P$ can be considered as a function of 
density $\rho$ only, thus
$P=P(\rho)$. 
As the characteristic oscillation periods
associated with inertial modes are in general
significantly  shorter than the global thermal timescale
we may adopt the approximation that 
 perturbations of the planet can be
 assumed to be adiabatic.  Then
the relation  
$P=P(\rho)$ holds during  perturbation as well 
 leading  to a barotropic equation of state.
In the  barotropic approximation the linearised 
Euler equations take the form (see PI)
\begin{equation}
{D^{2} {\bmth{ \xi}} \over Dt^{2}}+2{\bmth {\Omega}}\times {D \bmth
{\xi}\over Dt}=-\nabla W, \label{eq p3} 
\end{equation}
where
\begin{equation}
W=c_{s}^{2}\rho^{'}/\rho+\Psi^{int}+\Psi^{ext},  \label{eq p4}
\end{equation}
${\bmth{ \xi}}$ is the Lagrangian displacement vector,
$\rho^{'} $ is the
density perturbation, $c_{s}$ is the 
adiabatic sound speed, $\Psi^{int}$ is the  stellar  gravitational 
potential  arising from the  perturbations and  $\Psi^{ext}$ is 
an external forcing potential, say, the tidal potential in the problem of excitation 
of inertial waves by tides, see PI and IP.

\noindent The linearised continuity equation is    
\begin{equation}
\rho^{'}=-\nabla \cdot (\rho {\bmth {\xi}}).  \label{eq p6}
\end{equation}

Note that the centrifugal term is absent in equation  $(\ref{eq p3})$
being formally incorporated into the potential governing
the static equilibrium of the unperturbed star.
The convective derivative ${D\over Dt} \equiv {\partial \over \partial t}$
as there is no unperturbed motion in the rotating frame.

Although  incorporation of the  perturbation to the internal gravitational 
potential presents no principal difficulty, in this paper, for
simplicity we neglect it,
setting  $\Psi^{int}=0$. This procedure known as 'the
Cowling approximation' can be formally justified in the case when perturbations
of  small
spatial scale  in the WKBJ limit are considered. However,
it turns out that when  low frequency  inertial modes are considered 
the Cowling approximation has been found to  lead to results which are  
in qualitative and quantitative 
agreement with those obtained numerically for global modes obtained 
 with a proper treatment of perturbations to the gravitational potential (see below).
Therefore, we do not expect that the use of the Cowling
approximation can significantly influence our main conclusions.

Provided that the expressions for the density and sound speed 
are specified  for some unperturbed model of
the planet, the set of equations  $(\ref{eq p3}-\ref{eq p6})$ is complete.
Now we express the Lagrangian displacement vector and the density perturbation in terms
of  $W$ with help of equations $(\ref{eq p3})$ and $(\ref{eq p4})$,
and substitute the result into the continuity equation  from which we obtain
an equation for its Fourier transform in the form
\begin{equation} 
\sigma^{2}  {\bmth{A}} \tilde W_{m} -\sigma {\bmth{B}}
\tilde W_{m} -{\bmth{C}}\tilde W_{m} =
\sigma^{2}d{\rho \over c_{s}^{2}}
(\tilde \Psi^{ext}_{m}-\tilde W_{m}),  \label{eq 4}
\end{equation}
where $d=4\Omega^{2}-\sigma^{2}$, and
\begin{equation} 
{\bmth {A}}=-{1\over \varpi}{\partial \over \partial \varpi}
\left (\varpi \rho{\partial\over \partial \varpi}\right )-{\partial \over \partial z}\left (\rho
{\partial \over \partial z}\right )+
{m^{2} \rho \over \varpi^{2}},  \label{eq 5}
\end{equation}
\begin{equation} 
{\bmth {B}} =-{2m\Omega \over \varpi}{\partial \rho \over \partial \varpi}, 
\quad {\bmth {C}}=-4\Omega^{2}{\partial \over \partial z}\left(\rho {\partial
\over \partial z}\right).  \label{eq 6}
\end{equation}
It is very important to note that the operators ${\bmth {A }}$, 
${\bmth {B}}$ and ${\bmth {C}}$ are self-adjoint  when the inner product 
\begin{equation}
(W_{1}|W_{2})=\int_{V} dz \varpi d\varpi W^{*}_{1}W_{2},  \label{eq 7}
\end{equation} 
with $V$ denoting the volume of the star. Here these operators are assumed to act
on well behaved functions and the density is taken to vanish at the surface boundary.
 Also when  $m \ne 0$
${\bmth {A}}$ and ${\bmth {B}}$ are positive 
definite and  ${\bmth {C}}$ is non negative. 
When $m=0,$
 ${\bmth {A}}$ remains positive definite
 if  consideration is restricted to  the physically acceptable
 variations $W$ that conserve mass,
 this constraint eliminating the possibility that $W$ is constant.

When the Cowling approximation is adopted equation (\ref{eq 4}) fully
specifies solutions to the problem of forced linear perturbations of a
rotating barotropic planet. In the general case, a complete set of 
equations is described in PI.

\subsection{The anelastic approximation}\label{anelastic} 

\noindent When $\tilde \Psi^{ext}_{m} = 0,$ equation (\ref{eq 4})
leads to an eigenvalue  problem 
describing  the free oscillations of a rotating star in the form 
\begin{equation}
\sigma^{2}  {\bmth{A}} \tilde W_{m} -\sigma {\bmth{B}}
\tilde W_{m} -{\bmth{C}}\tilde W_{m} =
-\sigma^{2}d{\rho \over c_{s}^{2}}
\tilde W_{m},  \label{eq 41}
\end{equation}

Assuming that
rotation of the planet is relatively slow such that 
the angular velocity $\Omega \ll \Omega_{*}$, 
these may be classified as
$f$ or $p$ modes with eigenfrequencies such that $|\sigma| > \Omega_{*}$ 
 or inertial modes with eigenfrequencies $|\sigma| \sim \Omega$. 
The $f$ and $p$ modes exist in non rotating stars and can be
treated in a framework of perturbation theory taking advantage 
of the small parameter
$\alpha = |\Omega/\sigma|$
(see eg. Ivanov $\&$ Papaloizou 2004  and references therein).  

On the other hand  for inertial 
waves $\alpha$ is of order unity, 
 such a perturbation approach cannot be used. 
Since, in general, equation (\ref{eq 41}) is
rather complicated even for  numerical solution, in order to
make it more tractable
the so-called 'anelastic approximation'   has been frequently used (see 
eg. PP, Lockitch \& Friedman 1999
and Dintrans \& Ouyed  2001) for which the
right hand side of (\ref{eq 41}) is neglected. 
  
In order to justify this approximation we note that 
for eigenfunctions  that are 
 non singular everywhere in the planet,  
we can  crudely estimate the
derivatives entering equation  (\ref{eq 41}) as 
$|\nabla W| \sim kW/R_{*}$ and $|\nabla^{2} W| \sim k^{2}W/R^{2}_{*}$,
where the parameter $k > 1$. 
Consider  first the interior region of the
planet where we approximately have $c_{s}^{2} \sim
GM_{*}/R_{*}$.  
It follows from equation (\ref{eq 41}) that   
the left hand side  and the right hand side  
can be respectively  estimated as 
$\sim \Omega^{2}\rho k^{2}\tilde W_{m}/R_{*}^{2}, \quad {\rm and} \quad
\Omega^{4}\rho \tilde W_{m}/ c_{s}^{2}.$
The ratio of these is  of order 
\begin{equation} \hspace{6cm}  {\Omega^{2} R_{*}^{2}\over c_{s}^{2}k^{2}} \sim 
{\Omega^{2}\over\Omega_*^2 k^{2}} \ll 1.  \label{eqn3}\end{equation}
This estimate is, however, not valid near the boundary of the planet 
where $c_{s}^{2}\rightarrow 0$ 
and the  left hand side of the inequality
(\ref{eqn3}) diverges. 
However, in the same limit 
the terms containing the density gradient on the left
hand side of (\ref{eq 41})  will dominate  terms involving the 
second derivatives of $\tilde W_{n}$.   Thus in this limit 
the magnitude of the contribution from terms on the  left hand side 
of (\ref{eq 4}) may   be estimated
 to be
\begin{equation} \hspace{6cm} {\Omega^{2}}\left|{d\rho\over dr}\right| k\tilde W_{m}/
R_{*} \sim  \Omega^{2}\Omega_{*}^{2}{\rho k\tilde W_{m}/
c^{2}_{s}}, \label{eqn4}\end{equation}
where we remark that it follows from  hydrostatic
equilibrium  that close to the surface
 $|d\rho / dr|\sim \rho GM_{*}/(c_{s}^{2}R_{*}^{2}).$
Accordingly, when $r \rightarrow R_{*}$ the
ratio of the terms  on the right and left hand sides of equation (\ref{eq 41})
can be estimated as
\begin{equation} \hspace{6cm} \sim {\Omega^{2}\over\Omega_*^2 k} \ll 1 \label{eqn5}\end{equation}
From equations (\ref{eqn3}) and (\ref{eqn5}) it follows that when $\Omega \ll
\Omega_{*}$  the terms determining deviation from the 
anelastic approximation are  small compared to the leading terms
everywhere in the planet. Accordingly, in the slow rotation regime,
 we can use this approximation
to find the leading order solutions for eigenfrequencies and 
eigenfunctions  and then proceed to  regard the terms on the right hand side of 
(\ref{eq 41}) as
a perturbation.

The validity of the anelastic approximation in the context 
of the tidal excitation of inertial modes has been recently questioned
in a recent paper by Goodman \& Lackner (2009) on account of the divergence
of the terms on the right hand side of equation(\ref{eq 41}) as $r \rightarrow R_*$
although an actual demonstration of its failure was not given.
In fact the above discussion, which also applies to equation (\ref{eq 4})
as this differs only by the addition of a forcing term,
 indicates that these terms are never important
provided $\Omega/\Omega_*$ is sufficiently small.  This is to be expected
because as $\Omega$ is reduced, the structure of the modes remains unaffected
in the anelastic approximation whereas the radial width  of the region where
terms on the right hand side of equation(\ref{eq 41})
might become comparable to any other
terms shrinks to zero.

We also note that the vanishing of the normal velocity at the boundary 
in the anelastic approximation is correct in the limit $\Omega/\Omega_* \rightarrow 0$
as the ratio of the  horizontal to normal components there
 can be shown using the above arguments to also be on the order of   $(\Omega/\Omega_*)^2.$
Finally in PIN, 
we find by comparing the results of tidal forcing calculations
using  a spectral approach with the anelastic approximation, to those obtained
using direct numerical solution of the initial value problem, that it gives good results
even when $\Omega/\Omega_*$  is not very small.

\subsection{Self-adjoint formalism}

It was shown by PI and IP that both
quite generally and also  when the  anelastic approximation is
used  equations (\ref{eq 4}) and (\ref{eq 41})  can be brought
to the standard form  leading to an eigenvalue  problem for a  
self-adjoint operator. Here we describe the approach, which leads to the 
self-adjoint formulation of the problem in the anelastic approximation.

The self-adjoint and non negative
character of the operators ${\bmth {A}}$, ${\bmth{B}}$ and
 ${\bmth{C}}$  is made use of to formally introduce their square roots, eg. ${\bmth {A}}^{1/2}$,
defined by condition ${\bmth {A}}={\bmth {A}}^{1/2}
{\bmth {A}}^{1/2}$, etc. As is standard, the requirement of non negativity, makes the  definitions of these
square roots unique.  The positive definiteness of ${\bmth {A}}$  (see above discussion)
also allows
definition of the inverse of ${\bmth {A}}^{1/2}$, ${\bmth {A}}^{-1/2}$.

Let us consider a new generalised  two dimensional 
vector  $\vec Z$ with components 
such that $ \vec Z  =
(Z_1 , Z_2)$
 and the straightforward generalisation of the  inner product
given by equation (\ref{eq 7}). It is now easy to see that equation
(\ref{eq 4}) is equivalent to  
\begin{equation}
\sigma \vec Z ={\bmth {H}} \vec Z +\vec S,  \label{eq 12} 
\end{equation}
where
\begin{equation}
{\bmth {H}}=  
\left( \begin{array}{cc} {\bmth {A}}^{-1/2}{\bmth {B}}
{\bmth {A}}^{-1/2} & 
{\bmth {A}}^{-1/2} {\bmth {C}}^{1/2} \\
{\bmth {C}}^{1/2} {\bmth {A}}^{-1/2} & 0 \end{array}\right), \label{eq 13}
\end{equation} 
and the  vector $\vec S_{}$ has the components
\begin{equation} ({\bmth {A}}^{-1/2}\sigma^{2}d{\rho \over c_{s}^{2}}
(\tilde \Psi^{ext}_{m}-\tilde W_{m}), \quad 0) \label{eqn6}.\end{equation}
Note that as follows from (\ref{eq 12})  the relation between the components of $\vec Z$ and $\tilde W_{m}$
can be taken to be  given by  
\begin{equation}
Z_1  =  \sigma {\bmth {A}}^{1/2}  \tilde W_{m},  \qquad  Z_2 = {\bmth {C}}^{1/2}  \tilde W_{m}.
\label{eq 13n}
\end{equation} 
Since the off
diagonal elements in the matrix (\ref {eq 13}) are adjoint of each
other and the diagonal elements are self adjoint, 
it is clear that the operator  ${\bmth{ H}}$ is self-adjoint.
Equation (\ref{eq 12}) can be formally solved using the spectral
decomposition of  ${\bmth{ H}}.$ We now make a few remarks 
concerning the spectrum. 

\subsection{The  oscillation spectrum of a rotating fluid  contained within an axisymmetric domain}\label{sec2.5}
It has been known for many years (see eg. Greenspan 1968, Stewartson
$\&$ Rickard 1969)
that  the eigenvalue problem we consider is not well posed in the inertial mode range $-2\Omega < \sigma < 2\Omega.$  This is
because in this spectral range the eigenvalue equation (\ref{eq 41}) becomes a hyperbolic partial differential equation
with boundary conditions specified on the planet boundary. The form of the spectrum depends on the behaviour 
of the characteristics,  which correspond to localised inertial waves,  under successive reflections from the boundary.
Note that these reflections maintain a constant angle with the rotation axis rather than the normal to the boundary.
The situation was conveniently  summarised
by Wood (1977) (see also Fokin 1994a,b and references therein). There are three types of behaviour of the characteristic paths for frequencies in the inertial mode range.
They may all close  forming periodic trajectories, they may be ergodic, or there may be a finite number of periodic trajectories
that form attractors. The first two types of behaviour  are believed to be associated with discrete  normal modes while
the third type leads to wave attractors and a continuous spectrum.  The homogeneous sphere  within
a spherical or ellipsoidal boundary  exhibits the first two kinds of behaviour and has discrete  normal modes
which form a dense spectrum 
(eg. Bryan 1889)  
while the same system with a solid core has wave attractors (eg. Ogilvie \& Lin 2004, Ogilvie 2009).
Note the characteristics behave in the same way for all spheres or ellipsoids with a continuous density distribution
so that these should have normal modes. Note too that in the limit of very short wavelength modes
only the second derivative terms matter in equation (\ref{eq 41}) and the system becomes equivalent
to the two dimensional case studied by  Ralston (1973) and Schaeffer (1975). In that case the normal modes are associated
with the frequencies for which all  characteristic paths are periodic. They form a dense spectrum  and are infinitely degenerate.
From this discussion we expect the modes of a system with a continuous density distribution
to approach the same form as those of the homogeneous sphere, an aspect upon which we build our 
later WKBJ approach.

\subsubsection{Formal solution of (\ref{eq 12}) in the anelastic approximation}
From the above discussion we expect the normal modes for the cases of interest
 to form a discrete but dense
spectrum.
The anelastic approximation can be implemented by setting $\tilde W_{m}=0 $
in equation (\ref{eqn6}). In this case we can look for a
solution to  (\ref{eq 12}) in
the form 
\begin{equation} \vec Z=\sum_{k} \alpha_{k} \vec Z_{k}, \label{eq 13a}\end{equation}
where 
 $\vec Z_{k}$ are the real eigenfunctions of ${\bmth{ H}}$ satisfying
\begin{equation}  
\sigma_{k} \vec Z_{k} ={\bmth{ H}} \vec Z_{k},  \label{eq 14} \end{equation}
the associated necessarily real 
eigenfrequencies being  $\sigma_k.$ 

\noindent Substituting  (\ref{eq 14}) into  (\ref{eq 12}) we obtain
\begin{equation}  
\alpha_{k} ={<\vec Z_{k}| \vec S>\over <\vec Z_{k}|\vec
Z_{k}>(\sigma-\sigma_{k})}.  \label{eq 15}
\end{equation}
 The operator
${\bmth{ H}}$ induces the inner product and  associated orthogonality relation
for eigenfunctions  according to the rule
\begin{equation}<\vec Z_{k}| \vec Z_{l}>=\sigma_{k}\sigma_{l}(W_{k}| {\bmth {A}}
W_{l})+(W_{k}|{\bmth {C}} W_{l}) = \delta_{kl},\label{eqn7}\end{equation} 
where 
\begin{equation} W_{k}={\bmth
{C}}^{-1/2}Z_{2}^{k}=\sigma_{k}^{-1}{\bmth A}^{-1/2}Z_{1}^{k}. \label{eqn8}
\end{equation}
Using  (\ref{eqn6}) and  (\ref{eq 15})
we explicitly obtain
\begin{equation}
\tilde W_{m}=
\sigma^{2}d\sum_{k} {\sigma_{k}\over N_{k}(\sigma-\sigma_{k})} 
(W_{k}| {\rho \over c_{s}^{2}}\tilde \Psi^{ext}_{m}) W_{k},  \label{eq 16}  
\end{equation}
where 
\begin{equation}N_{k}=\sigma^{2}_{k}(W_{k}| {\bmth {A}}
W_{k})+(W_{k}|{\bmth {C}} W_{k})\label{eq 16a}\end{equation} 
is the norm. The decomposition (\ref{eq 13a}) should be valid for any
vector $\vec F$ with components $(F,0)$, where $F$ is any function of
the spatial coordinates. The second component of this equality
shows that in order for this to be valid an identity 
\begin{equation}\sum_{k} {<\vec F|\vec Z_{k}> \over N_{k}}Z_{2}^{k}=0 \end{equation}
must be hold (IP). This identity allows us to represent the relation (\ref{eq 16})
in a different form (PI):
\begin{equation}
\tilde W_{m}=
\sigma d\sum_{k} {\sigma_{k}^{2}\over N_{k}(\sigma-\sigma_{k})} 
(W_{k}| {\rho \over c_{s}^{2}}\tilde \Psi^{ext}_{m}) W_{k}.  \label{eq 16b}  
\end{equation}
Note that in response problems such as the problem of excitation of the
inertial waves during the periastron flyby, in order to take
account of causality issues correctly when extending to the complex
$\sigma$ plane, one should add a small
imaginary part in the resonance denominator in  (\ref{eq 16b})
according to the Landau prescription: 
$(\sigma-\sigma_{k}) \rightarrow (\sigma +i\nu-\sigma_{k})$, where
$\nu > 0$ is a small real quantity.

\subsubsection{Corrections to the anelastic approximation}

When external forces are absent and the potential $\tilde
\Psi^{ext}_{m}$ is set to zero, equation  (\ref{eq 4}) (or,
alternatively, equation (\ref{eq 12}))  defines   the full eigenvalue
problem. Under very general assumptions it was shown by IP that 
this problem can  be formally solved in an analogous manner. 
However, 
it is rather difficult to use the general expressions obtained by IP
without making further approximations.
Here we note that, given that the spectrum is discrete, we may find conditions satisfied
by the eigenfunctions and eigenvalues by  replacing  $\tilde
\Psi^{ext}_{m}$   by  $-\tilde
W_{m}$ in equations (\ref{eq 16}) and (\ref{eq 16b}).
These conditions relate any eigenfunction, now equated to  $\tilde
W_{m}$  and its  associated eigenvalue $\sigma$ to the eigenfunctions and eigenvalues
of the anelastic problem.
Proceeding in this way we go on to form 
the quantity
\begin{equation}\sigma\sigma_{l}(W_{l}| {\bmth {A}}
\tilde W_{m})+(W_{l}|{\bmth {C}} \tilde W_{m}) = - {\sigma^2 d \sigma_{l}\over (\sigma-\sigma_{l})}
(W_{l}| {\rho \over c_{s}^{2}}\tilde W_{m}), \label{eqn16c}\end{equation}
where $W_{l}$ is an anelastic eigenfunction and we have made use of the orthogonality relation
(\ref{eqn7}).

As argued in section \ref{anelastic}, the quantity on the right hand side
can be regarded as a perturbation where the small parameter is $\epsilon = (\Omega/\Omega_*)^2.$
Provided an eigenfunction can be identified as $\epsilon \rightarrow 0$ and is non degenerate
with $\tilde W_{m}\rightarrow W_{l},$ it follows from (\ref{eqn16c}) that in this limit

\begin{equation}
(\sigma-\sigma_{l})
 = -{ d_l\sigma^3_{l}\over N_{l}}
(W_{l}| {\rho \over c_{s}^{2}} W_{l}), \label{eqn16d}\end{equation}
where $d_{0}=4\Omega^{2}-\sigma_{l}^{2}$.

The spectrum of inertial modes is 
dense. This may lead to a potential difficulties in identifying and following
modes as parameters change as we discussed above. However, it is possible to argue that
this problem can be alleviated for large scale global modes by for example modifying
the eigenvalue problem by adding terms that have a very small effect on the global modes
but spectrally separate close by short scale  modes. Dintrans and Ouyed (2001) adopt such a procedure
by adding a viscosity and this enables them to 
identify and follow global modes. Note   that a similar situation 
 would result if conservative high order derivative
terms were added that preserved the self-adjoint form of the problem.
Numerical  work presented below and in PIN also confirms that global modes
have a clear identity and can be followed as parameters change provided that the angular frequency is
sufficiently small. Thus we both expect and verify the validity of the expression (\ref{eqn16d}) in this limit.
 For larger values of $\Omega $ one should take into account a possibility of mixing
between two neighbouring large scale global modes to explain results of numerical calculations, see PIN.
 In this case expression (\ref{eqn16d}) should be modified in an appropriate way.


\subsubsection{Eigenvalues corresponding to opposite signs of $m$}

In the next section we  find solutions of the eigenvalue
problem in the WKBJ approximation. It will be shown that  
the corresponding eigenvalues and eigenmodes are 
independent  of the  sign of  $m$  to 
 two leading orders. This is explained by the fact that  to that order solutions 
are determined only by operators containing second and first
derivatives in equation (\ref{eq 4}). On the other hand 
it follows from the same equation that the only dependence on sign of $m$ is 
determined by the operator ${\bmth {B}}$ which does not contain any
derivatives of $W$.   

In order to find  the first correction to the WKBJ eigenfrequencies
that depends on sign of $m,$ we  treat the operator  ${\bmth {B}}$ as a perturbation.
This leads to a change in the eigenfrequency that can be found by using 
the same formalism that lead to equation (\ref{eqn16d}) but  then simply replacing
$\sigma_l d_l \rho  W_{l} / c_{s}^{2}$ in that equation by $- {\bmth B} W_{l}.$
Equation (\ref{eqn16d}) then gives
 \begin{equation}
 \sigma-\sigma_{l} =-2m\Omega \sigma_l^{2}\left(\int d\varpi dz {\partial \rho
  \over \partial \varpi} W_{l}^{2}\right )/N_{l}.  
\label{eqnn2}
\end{equation}
Note that since ${\partial \rho
\over \partial \varpi} < 0$ it follows from equation (\ref{eqnn2})
that when $\Omega > 0$ 
the sign of
$\sigma - \sigma_{l}$ is proportional to the sign of $m$.

\subsection{A form of equation (\ref{eq 4}) valid for a spherical planet}\label{sec2.6}     

In what follows we assume that an object experiencing
tidal interactions can be approximated as having a  spherically
symmetric structure. In this case it is
appropriate to use another form of (\ref{eq
4}) with $\tilde \Psi^{ext}_{m}=0$, which is especially convenient for an
analysis of WKBJ solutions. We can obtain this from
(\ref{eq 4}) 
using the fact that for a spherical star ${\partial \over \partial \varpi}\rho
={\varpi \over r} {d\over dr}\rho $ and ${\partial \over \partial z}\rho
={z \over r} {d\over dr}\rho $. We  obtain
\begin{equation}
\left[\sigma^{2}\Delta  -4\Omega^{2}{\partial^{2} \over \partial z^{2}}\right] W
={1\over rH}\left(\left[\sigma^{2} \varpi {\partial W \over \partial \varpi}
-d z {\partial W \over \partial z }\right]  -\left[2m\sigma \Omega
-{d (\sigma/ \Omega_{K}(r))^{2}}\right]W\right),   
\label{eqn12}
\end{equation}
where we set, for simplicity, $W\equiv \tilde W_{m}$, 
$\Delta $ is the Laplace operator, $H=-{d r\over d \ln\rho}$ is
a characteristic density scale height and $\Omega_{K}(r)=\sqrt
{GM(r)\over r^{3}}$, where $M(r)$  is the mass enclosed within a
radius $r$. Note that we use the 
hydrostatic balance equation
$-c_{s}^{2}{d\rho \over dr}={GM(r)\over r^{3}}$ to obtain equation 
(\ref{eqn12}) from equation (\ref{eq 4}). The last term in the second 
square braces on the right hand side describes correction to the 
anelastic approximation. It is discarded when the WKBJ approximation 
is used. 

\subsubsection{Pseudo-spheroidal coordinates}  

 When the density approaches a constant value,
 $H$  tends to infinity and  the right hand side of equation (\ref{eqn12})
vanishes. In this case it describes an
incompressible fluid, see eg. Greenspan (1968).
It was shown by Bryan (1889) that in this case this equation is
separable in special 'pseudo-spheroidal' orthogonal coordinates defined
by  the relations
\begin{equation}\varpi =R_{*}\sqrt{{(1-x_{1}^{2})(1-x_{2}^{2})\over
(1-\mu^{2})}}, \quad z={R_*x_{1}x_{2}\over \mu}, {\rm  \ \ where \ \ the\ \ constant\ \ parameter\ \ }
  \mu ={\sigma \over 2\Omega}. \label{eqn13}           
\end{equation}                 
Since the governing equations are  invariant to the mapping $(\sigma, m)
\rightarrow (-\sigma, -m)$, without loss of generality we assume 
from now on that $\mu > 0$ for all modes while  $m$ can 
have either sign. 
Also, from equation (\ref{eqn12}) it follows that the
modes should be either even or odd with respect to the reflection in the equatorial plane
$z\rightarrow -z$. Therefore, it is sufficient to consider only the
upper hemisphere $(z >0, \sqrt{\varpi^{2}+z^{2}} \le R_{*})$. In this
region we can assume that the variables $x_{1}$ and $x_{2}$  are contained within the
intervals $[\mu, 1]$ and $[0,\mu]$, respectively. A detailed
description of this coordinate system can be found in eg. Arras et
al (2003), Wu (2005)a.  
   
Using  the new variables equation (\ref{eqn12})  takes the form
\begin{equation}
(\hat D_{2} -\hat D_{1})W={1\over \mu^{2}(1-\mu^{2})Hr} 
\left( \hat AW
-(x_{1}^{2}-x_{2}^{2})(m\mu-4\mu^{2}(1-\mu^{2})({\Omega /\Omega_{K}(r)})^{2})W\right)
\label{eqn15} 
\end{equation} 
where
\begin{equation}
\hat D_{i}= -{d\over d x_{i}}(1-x_{i}^{2}){d\over d x_{i}}
+{m^{2}\over (1-x_{i}^{2})},
\label{eqn16} 
\end{equation} 
\begin{equation}
\hat A = x_{1}(1-x_{1}^{2})(x_{2}^{2}-\mu^{2}){\partial \over \partial x_{1}}
-x_{2}(1-x_{2}^{2})(x_{1}^{2}-\mu^{2}){\partial \over \partial
x_{2}},
\label{eqn17} 
\end{equation}
and the quantities $r$, $H$, $\Omega_{*}(r)$ are understood  to be 
functions of the variables $x_{1}$ and $x_{2}$. It is easy to see that 
the eigenfunctions of the operators $\hat D_{i}$ are the associated Legendre functions, 
$P_{\nu}^{m}(x_{i})$,  and we have
\begin{equation} D_{i}P_{\nu}^{m}(x_{i})=\lambda^{2}P_{\nu}^{m}(x_{i}), 
\label{eqn18} 
\end{equation}
where $\lambda=\sqrt{\nu(\nu+1)}$. Let us stress that as the domains
of $x_1$ and $x_2$ are not $[-1,1],$ $\nu $ is not
necessarily an integer.  

In some important cases equation (\ref{eqn15}) is separable.
Firstly, when the gas is incompressible,  the right  hand 
side of (\ref{eqn15}) is zero. In this case it follows from
equation (\ref{eqn18}) that the solution can be represented as
product of two associated Legendre functions. 

Secondly, as  was
mentioned by Arras et al (2003) and later  explored in detail by Wu (2005)a,b
 when equation (\ref{eqn15}) is considered in the anelastic
approximation it is separable for planetary models  
with  density profiles of the form
\begin{equation}\rho=C\left(1-\left({r\over R_{*}}\right)^{2}\right)^{\beta}, \label{eqn19}\end{equation}
where $C$ is a constant.
These models include the incompressible one which corresponds 
to $\beta=0$. It was also noted by Arras (2003) and  Wu (2005)a,b that  for
polytropic models with equation of state
\begin{equation}
\hspace{6cm} p=k\rho^{\gamma},
\label{eqn20} 
\end{equation}
close to surface the density distribution has the form 
\begin{equation}\hspace{6cm} \rho=Dx^{n},    \label{eqn21}          \end{equation}
where $x=1-r/R_{*}$, $n=1/ (\gamma -1) $ is the 
polytropic index,  and $D$ is a constant. In the asymptotic 
limit $r\rightarrow R_{*}$ this
expression coincides with what is obtained from equation
(\ref{eqn19}) with  $\beta=n$. This proves that when  polytropic
models are considered equation (\ref{eqn15}) is separable in a plane 
parallel approximation often adopted  close to the
surface. Here we would like to note that this is valid even when the 
anelastic approximation is relaxed since close to the surface we
have $\Omega_{K}(r)\approx  \Omega_{*}=const$ and the additional term appearing 
in this case in the braces on the right hand side of (\ref{eqn15}) has the same  spatial
structure of a term already present in the anelastic approximation.

\section{WKBJ solutions for the normal modes }\label{sec3}

In general equation (\ref{eqn15}) should be solved numerically. We
can, however, look for analytical solutions to (\ref{eqn15}) in the WKBJ
approximation assuming that solutions are fast oscillating functions
in the planet's interior. The first and second derivatives of these
functions are assumed to be proportional to first and second power of
a large parameter $\lambda $, the value of which is specified below.
This problem has been analysed before by Arras (2003) and Wu (2005)a who 
obtained  expressions for eigenvalues and eigenfunctions for the 
problem of free oscillations in the inertial mode
spectral range. Here we revisit  the problem,   
taking into account  terms that appear at the next order
 in an asymptotic expansion of the
quantities of interest in a  power series in $1/\lambda.$  This
will allow us to obtain analytic expressions which 
agree with numerical results,  even   for the rather small 
values of $\lambda $, appropriate  to  global modes 
(see below).

\subsection{Natural units}

In what follows in order to simplify notation we express all
dimensional quantities in natural units.  
These are such that the spatial coordinates, density, angular velocity and 
sound speed are expressed in units of $R_{*}$, the   mean density 
${\bar \rho}=3M_{*}/ (4\pi R_{*}^{3})$,$\sqrt {GM_{*}/
R_{*}^3}$ and $\sqrt {GM_{*}/
R_{*}}$ respectively. All other quantities of interest are expressed in 
terms of powers of these basic units.

\subsection{WKBJ solutions}

It is easy to find from either (\ref{eqn12}) or  (\ref{eqn15}) that in the planet's interior
far from the rotational axis, the WKBJ solution should have the form
\begin{equation} 
W_{WKBJ}={1\over \sqrt{\rho \varpi}}(C_{1}e^{i\lambda \phi_{+}(u_{+})} +
C_{2}e^{i\lambda \phi_{-}(u_{-})}+cc) +O({1\over \lambda}), 
\label{eqn22}\end{equation}
where  
$\phi_{\pm}(u_{\pm})$ are arbitrary functions of 
\begin{equation} 
u_{\pm}=\varpi \pm {\mu \over \sqrt{1-\mu^{2}}}z, 
\label{eqn23}\end{equation}
the constancy of which defines the characteristics of equation (\ref{eqn15}). 
Acceptable  forms for the  functions $\phi_{\pm}$ have to  be determined by matching the
solution (\ref{eqn22}) to approximate solutions valid near the surface  boundary
and near the rotational axis. It turns out that this matching is
possible if the WKBJ solution has the form
\begin{equation} 
W_{WKBJ}={1\over \sqrt{\rho \varpi}}(\cos(\lambda y_{1}+\phi_{1})\cos(\lambda
y_{2} + \phi_{2})),
\label{eqn24}\end{equation}
where $y_{1,2}=\arccos( x_{1,2})$ and $\phi_{1,2}$ are  constants to
be determined. One can readily check with help of the coordinate transformations
(\ref{eqn13}) that this form agrees with the general expression 
(\ref{eqn22}), see equation (\ref{eqn50}) below. 
Since $y_{1,2}$ are multivalued functions of $x_{1,2}$ 
we should specify a one-valued branch of these. Taking into account
that our calculations 
will be done for positive values of $x_{1,2}$, we assume below that values
of $y_{1,2}$ are in the range $[0,\pi/2]$.

For simplicity, in the main text we are going to consider the modes even with respect to
reflection $z\rightarrow -z$, called hereafter `the even modes'. 
For example, such modes are excited by tidal interactions since
tidal potential is an even function of $z$. The case of the modes odd with respect to
this reflection ('the odd modes') 
can be dealt with in a similar way. This case is considered in Appendix A.

From equation (\ref{eqn13}) it follows that reflection
of the coordinate $z$ leads to the reflection of  the coordinate $x_2$ such that
$x_{2}\rightarrow -x_{2}$,  while the coordinate $y_{2}$ changes 
according to the rule $y_{2}\rightarrow \pi -y_{2}.$  We readily find that
(\ref{eqn24}) is  unchanged under  this
transformation provided  the phase 
\begin{equation} \phi_{2}=-\pi\lambda/2, \label{eqnn24}\end{equation} 
(see also eg. Wu 2005a). We remark that the same result is obtained
by requiring that the derivative of (\ref{eqn24})  with respect to $y_2$ vanish 
on the equator where  $y_2=\pi/2.$

\subsection{Matching near the rotation axis}
In the WKBJ approximation sufficiently far from 
the rotational axis all 
terms proportional to $W$ give small corrections to the
solution (\ref{eqn22}) and are formally discarded.   However, when $\varpi \rightarrow
0$ and, accordingly, $x_{1}\rightarrow
1$,  it follows from equation (\ref{eqn16}) that the term proportional
to $m^{2}W$ in the expression for the operator $\hat D_{1}$
diverges in this limit and should be retained. When this is done
the phase $\phi_{1}$ can be found from condition of regularity of $W$ close to
the rotation axis $\varpi=0$.

We begin by using the WKBJ solution already found to develop  an approximate  expression for
$W$  that is  appropriate for small
values of $\delta = 1-x_{1}$ and which can be  matched at large distances from
the rotation axis. 
An appropriate expression for $W$ which  can be matched to the
correct WKBJ limit sufficiently far from the rotational axis is
\begin{equation}
W\propto {1\over \sqrt{\rho} {(1-x_{2}^{2})^{1/4}}}\cos \lambda (y_{2}-\pi/2)W_{a}(\delta),
\label{eqn25} 
\end{equation}     
where we take into account that the factor $\sqrt{\varpi}$ entering
(\ref{eqn22}) is proportional to the product
$(1-x_{1}^{2})^{1/4}(1-x_{2}^{2})^{1/4}\propto
\delta^{1/4}(1-x_{2}^{2})^{1/4}$, see equation (\ref{eqn13}), 
and the factor $\delta^{1/4}$ is
formally incorporated in the definition of $W_{a}(\delta)$ which is to be found by
imposing the condition of regularity on the rotation axis.
 In order to do this we obtain an equation for $W_{a}$   
from  equation (\ref{eqn12}) (or (\ref{eqn15})) that retains  terms containing the
derivatives and terms that potentially diverge in the limit $\varpi \rightarrow
0$ while other terms can be discarded.

 From equations
(\ref{eqn12}) and (\ref{eqn15}) it follows that
$W_{a}(\delta )$ satisfies 
equation (\ref{eqn18}) in the limit of small $\delta $
\begin{equation}   
\delta^{2}{d^{2}\over d\delta^{2}}W_{a}+\delta {d\over d\delta}W_{a} +({\lambda^{2}\over
  2}\delta -{m^{2}\over 4})W_{a}=0.
\label{eqn26} 
\end{equation}
The solution to (\ref{eqn26}) regular at the point $\delta=0$ can be
expressed in terms of the Bessel function
\begin{equation}  
W_a\propto J_{|m|}(\sqrt{2}\lambda \delta^{1/2}),
\label{eqn27} 
\end{equation}
where we assume from now on that $\lambda $ is positive\footnote{In
the approximation we  consider the final expressions are independent
of the change of sign of $\lambda$.}.  
In the limit of large $(\lambda \delta )$  the asymptotic form of the expression (\ref{eqn27}) 
is
\begin{equation}   
W_a\propto \delta^{-1/4}\cos\left (\sqrt{2}\lambda \delta^{1/2} -|m|{\pi \over 2}-{\pi\over 4}\right).
\label{eqn28} 
\end{equation}
It is easy to see that when $\delta $ is small
$y_{1}\approx
\sqrt{2\delta}$. Therefore, from equations (\ref{eqn25}) and
(\ref{eqn28}) it follows that the solution has the required form
(\ref{eqn24}) provided that
\begin{equation} 
\phi_{1}=-|m|{\pi\over 2}-{\pi\over 4},
\label{eqn29} 
\end{equation}
and we have, accordingly,
\begin{equation} 
W_{WKBJ}={1\over \sqrt{\rho \varpi}}F, \quad F=\cos\left(\lambda y_{1}-|m|{\pi\over
2}-{\pi\over 4}\right )\cos\left (\lambda y_{2} -\lambda{\pi\over 2}\right).
\label{eqn30} 
\end{equation}
Note that the phase (\ref{eqn29}), which can in fact be verified
with reference to the incompressible sphere,  differs from that  given in
Arras et al (2003) and Wu (2005)a. This disagreement is due to an
oversimplified treatment of the WKBJ solution close to the rotational axis in
these papers.  

\subsection{  Matching at the planet surface } \label{Surfmatch}
The eigenvalues appropriate to  the problem of free oscillations can be found by
matching the  solution (\ref{eqn30}) to approximate solutions valid near
the surface of the planet. 
In  pseudo-spheroidal coordinates
(\ref{eqn13}) the  equation determining the upper hemispherical surface of the planet $(r=R_{*}, z>0)$ 
has  two branches: 1) $x_{1}=\mu, 0 < x_{2} < \mu$ and
 2) $\mu < x_{1} < 1, x_{2}=\mu$.  In order to simultaneously consider solutions to
equation (\ref{eqn15}) that can be  close to either  of these  branches,  we introduce
two new coordinates  $\delta_{i}, $  with $i=1$ corresponding
to the first branch and $i=2$ corresponding to the second branch,  that are 
defined by the  relation 
\begin{equation} x_{i}=\mu \pm \delta_{i},
\label{eqn30a}\end{equation}
where the sign $+$ ($-$) corresponds to the 1st (2nd) branch,
and assume later on that  the $\delta_{i}$ are  small.

The form of the solutions close to the surface depends on  the density
profile. In what follows we the consider the planet models with a 
polytropic equation of state for which the density profile close to
the surface is given by equation (\ref{eqn21}). The variable
$x=1-r/R_{*}$ entering equation (\ref{eqn21}) can be expressed through 
$x_{j}$ and $\delta_{i}$ as
\begin{equation}
x=\mp {(x_{j}^{2}-\mu^{2})\delta_{i}\over \mu (1-\mu^{2})},
\label{eqn31} 
\end{equation} 
where we assume from now on that the upper (lower) sign corresponds to
the 1st (2nd) branch and the index  $i$  takes on the values $1$  ( first branch) and $2$ (2nd  branch)
with $j\ne i.$ 

We now look for solutions close to the surface that have  $\lambda \delta_j$ large but $\lambda \delta_i$ small.
This is possible because in the WKBJ theory $\lambda$ is a large parameter.
The domain for which $\lambda\delta_i$ is small for both $i=1$ and $i=2$
is called the critical latitude domain and will be considered separately below.
Solutions valid in all of these domains must match correctly
on to a solution of the form (\ref{eqn30})  in order to produce a valid eigenfunction.

Using equations (\ref{eqn21}) and (\ref{eqn24}) we can look for
a solution close to the surface in the form 
\begin{equation}
W\propto |x_{j}^{2}-\mu^{2}|^{-n/2}(1-x_{j}^{2})^{-1/4}\cos (\lambda
y_{j} +\phi_{j})W_{i}(\delta_{i}),
\label{eqn32} 
\end{equation} 
where we also use equation (\ref{eqn13}) in order to express the
factor $\varpi^{-1/2}$ in terms $x_{j}$ setting
$\delta_{i}=0$ there. Substituting this expression in equation
(\ref{eqn15}) and taking the limit $\delta_{i}\rightarrow 0$ we obtain
\begin{equation}
\delta {d^{2}\over d\delta^{2}}W+n{d\over d\delta }W
+ {1\over (1-\mu^{2})}(\lambda^{2}\delta \pm nm_{*})W=0, 
\label{eqn33} 
\end{equation} 
where, for simplicity, we omit the index $(i)$ in the quantities
$\delta_{i}$ and $W_{i}$, and
\begin{equation}
m_{*} =m-4\mu(1-\mu^{2})\frac{\Omega^{2}}{\Omega_*^2}.
\label{eqn34} 
\end{equation} 
We recall that the term proportional to $\Omega /\Omega_{*}$ 
 gives the  correction to the anelastic
approximation.  Since in the low frequency limit  $\Omega/\Omega_* $
 is assumed to be much smaller than
unity, this term is small and we approximately have $m_{*} \approx
m$. 

Equation (\ref{eqn33}) can be brought into a standard form by  the   change of
variables
\begin{equation}
\zeta=2i\kappa \delta, \quad \Psi= e^{i\kappa \delta}W=e^{\zeta/2}W, \  \  {\rm where}  \  \  \kappa={\lambda \over \sqrt{1-\mu^{2}}}.  
\label{eqn35} 
\end{equation} 
 Adopting these we obtain 
\begin{equation}
\zeta \Psi^{\prime \prime}+(n-\zeta)\Psi^{\prime}-{1\over 2}(n\pm i\chi)\Psi=0,
\label{eqn36} 
\end{equation} 
where a prime  denotes  differentiation with respect to $\zeta$ and
$\chi=nm_{*}/ (\lambda \sqrt{1-\mu^{2}}).$
This is the confluent hyper-geometric equation. Its solution  that is regular at
the surface is expressed in terms of the confluent hyper-geometric function
$\Phi(a,b,z)$ as
\begin{equation}
W\propto e^{-\zeta/2}\Phi\left ((n\pm i\chi)/2, n, \zeta\right ).
\label{eqn37} 
\end{equation} 
Note that this solution is similar to solutions of the
Schrodinger equation with the 
Coulomb potential describing wave functions belonging to continuous part of 
its spectrum, ( see eg. Landau $\&$ Lifshitz 1977).

In the limit of $|\kappa \delta| \gg 1$ we obtain from (\ref{eqn37})
\begin{equation}
W\propto \zeta^{-n/2}\left ({1\over \Gamma ({1\over 2}(n\mp i\chi))}\exp
\left( i\left(-{|\zeta|\over 2}\mp {\chi \over 2}\ln|\zeta|+{\pi n\over 4}\right)\right)+cc\right),
\label{eqn39} 
\end{equation} 
where $G(z)$ is the gamma function. 
Since the quantity $\chi $ is assumed to be small we can approximately
write
\begin{equation}
{1/ \Gamma ({1\over 2}(n\mp i\chi))}\approx 
(1\pm i\psi({n\over 2}){\chi \over 2})/ \Gamma
  ({n\over 2})   \approx \exp\left(i\psi({n\over 2}){\chi \over 2}\right)/ \Gamma ({n\over 2}),
\label{eqn40} 
\end{equation} 
where $\psi(z)\equiv {d\over dz}\Gamma(z)$ is the psi function. In the
same approximation equation (\ref{eqn39}) can be rewritten in the form
\begin{equation}
W\propto \zeta^{-n/2}\cos
\left ({|\zeta|\over 2}\pm {\chi \over 2}\left(\ln|\zeta|-\psi({n\over 2})\right)-{\pi n\over 4}\right ).
\label{eqn41} 
\end{equation} 

After substituting the result expressed by equation (\ref{eqn41})
 into (\ref{eqn32}) the
resulting expression should be of the general` form (given by \ref{eqn24})
evaluated close to the surface. This, however, cannot be realised
on account of the
presence of the factor $\ln|\zeta|$ in (\ref{eqn41}). This term, having
a coordinate dependence  
of order of  $1/\lambda$ after removing a constant phase
would  formally require terms  of that
order that are   not accounted for in the expressions (\ref{eqn22})
and (\ref{eqn24})
to enable matching, therefore to the order we are currently
working,  it is discarded.
Since only this term  depends on the sign
of $m$ and  on the correction to the anelastic approximation, 
both dependencies are absent in the resulting approximation. 

Another way of obtaining solutions to (\ref{eqn33}) compatible with the form
(\ref{eqn24}) inside the planet is to set to zero the small quantity
$\chi \propto 1/\lambda $ in equation (\ref{eqn37}). In this case the
solution can be expressed in terms of a Bessel function such that
\footnote{In order to obtain equation (\ref{eqn38}) we use the
relations $\Phi(a,b,z)=e^{z}\Phi(b-a,b,-z)$ and $\Phi(a,2a,2z)\propto
z^{1/2-a}e^{z}J_{a-1/2}(iz)$, see eg. Gradshteyn $\&$ Ryzhik 2000, pp
1013, 1014.}
\begin{equation}
W(\delta) \propto \delta^{(1-n)/2}J_{1/2(n-1)}(\kappa \delta).
\label{eqn38} 
\end{equation}
Note that this expression is equivalent to (\ref{eqn37})  when the anelastic approximation is adopted and
$m=0$. When $\kappa \delta \rightarrow \infty$ we get 
\begin{equation}
W_{i}\propto \delta_{i}^{-n/2}\cos
\left({\lambda \over \sqrt{1-\mu^{2}}} \delta_{i}-{\pi n\over 4}\right),
\label{eqn42} 
\end{equation} 
where we the index $(i)$ has been
restored and we use the explicit expression for $\kappa $. 
Substituting (\ref{eqn42}) into equation (\ref{eqn32}),
taking into account that the factor
$(|x_{j}^{2}-\mu^{2}|\delta_{i})^{-n/2} \propto \sqrt \rho $, and that
close to the surface we have
\begin{equation}
y_{i}\approx \arccos \mu \mp {\delta_{i}\over \sqrt{1-\mu^{2}}}.        
\label{eqn43} 
\end{equation}

\subsection{Determination of the eigenfrequencies}\label{sec3.5}
 
It  can now  be seen that the expression (\ref{eqn32}) has the required form
(\ref{eqn24}) provided that the phases $\phi_{i}$  satisfy
appropriate appropriate conditions. However,  these phases have  already
been determined from the requirements
of regularity on the rotation axis
and symmetry with respect to reflection in the equatorial plane
and are accordingly specified through  equation (\ref{eqn30})
which equation(\ref{eqn32}) must match.

It is readily seen that  the expressions
(\ref{eqn30}) and (\ref{eqn32}) can be compatible only for 
particular   choices of $\lambda $ and $\mu$. These compatibility
conditions determine the eigenspectrum of the problem in the WKBJ 
approximation. They are easily found 
 from equations (\ref{eqn30}), (\ref{eqn42}) and
(\ref{eqn43})to be given by 

\begin{equation} 
{\pi \over 4}n+\pi k_{1}=\lambda \arccos
(\mu)-{\pi\over 2}|m|-{\pi \over 4} \quad {\rm and} \quad
{\pi \over 4}n+\pi k_{2}=\lambda \left({\pi \over 2}-\arccos
(\mu)\right).
\label{eqn44} 
\end{equation}  
 Here $k_{1}$ and $k_{2}$ are positive or negative
 integers that  must be chosen in 
a way which ensures that the angle $\arccos(\mu) $ belongs to the branch
 $0 < \arccos(\mu) < {\pi\over 2}$.

 Adding the above relations
 we  obtain
\begin{equation} \lambda =2l+n+|m|+{1\over 2}, \label{eqn45}\end{equation}
where $l=k_{1}+k_{2}$. Substituting (\ref{eqn45}) into the first expression 
in (\ref{eqn44}) we obtain an expression for the eigenfrequency
\begin{equation} 
\mu =\frac{\sigma}{2\Omega}= \cos\left( \pi {k+{|m|/ 2}+(n+1)/4\over \lambda}\right),     
\label{eqn46} 
\end{equation} 
where we set $k\equiv k_{1}$ from now on.

As shown in Appendix A the modes with different symmetry with respect to
reflection $z \rightarrow -z$ (both the 'even' and the `odd' modes) can be described
by the same expression (\ref{eqn46}) provided that the expression for $\lambda $ changes 
to
\begin{equation}
\lambda=p+n+|m|+{1\over 2},
\label{eqn45n}
\end{equation}
where the integer $p$ is even for the even modes while for those with odd symmetry $p$ is odd.

For the WKBJ approximation to be valid $\lambda$ should be large, and,
accordingly, $l \gg 1$. We would like, however, to consider all
values of $l$ and $k$ allowed by our assumption that
$\lambda $ is positive and $\arccos(\mu) $ belongs to the interval 
$(0,\pi/2)$. 
These conditions imply that $l$ is positive and lead to inequality:
\begin{equation} 
-\left[{(1+n)\over 4}+{|m|\over 2}\right] \le k \le \left[l+{n\over 4}\right],
\label{eqn47} 
\end{equation}
where $[Q]$ means that integer part of $Q$ is taken. 

When $l $ and $k $ are sufficiently large one may neglect other
quantities in the argument of  the 
cosine in  equation (\ref{eqn46}). In this case 
one gets $\mu=\cos (\pi k/(2l))$ - an expression obtained in 
previous papers (see Arras 2003 and Wu 2005a). One may also consider the
limit of  an incompressible fluid by setting $n=0$ in 
(\ref{eqn46}). In this case the expression (\ref{eqn46}) gives
 the  correct asymptotic  eigenfrequencies  appropriate to  the high order  
modes of pulsation of an incompressible fluid in a rotating 
spherical container, see Appendix B for details.

\subsection{A general expression for eigenfunctions close to
the surface of the planet}\label{surface}
The purpose  here is to establish an
expression for $W$ that is  approximately valid in the whole region
close to the surface  where the separation of
variables is possible and the eigenfunction
can be written as the product
of functions of $x_1$ alone  and $x_2$ alone.
Also, this expression should
approach the $WKBJ$ expression (\ref{eqn30}) in the
limit of sufficiently large $\lambda x$ in order to have the norm  that
will be given
by equation (\ref{eqn57}) below.

Close to the planet  surface we have
 $x\equiv 1-r \ll 1$, the  density
profile can be represented in the form (\ref{eqn21}), and  equation (\ref{eq
4})  becomes separable in the pseudo-spheroidal coordinates $(x_{1},x_{2})$ 
(Arras et al 2003, Wu
2005a). As described already in section \ref{Surfmatch}, in these coordinates
 the  region close to the surface is described by two branches
 $ x_{i}=\mu \pm \delta_{i}, \ \ (i=1,2)  $ 
see equation~(\ref{eqn30a}).  We denote these branches as the  $(+)$ branch for which $i=1$
 and  the  $(-)$ branch for which $i=2,$  respectively. 

 When one of the coordinates,
 $x_{j},$ say,  is sufficiently far from the value $x_{j}=\mu$ the 
eigenfunction is proportional to the expression given by equation 
(\ref{eqn32}). In practice the requirement that $x_j$ is far  from $\mu$ is that  $|\lambda \delta_j|$ be large.
When this parameter is  of order unity or less, $x_j$ is considered to be close to $\mu.$
When both coordinates are close to $\mu$ in this sense,  the eigenfunction is proportional, 
with,  in the limit of large $\lambda,$ 
proportionality factor being slowly varying,  to the product
$W_{1}(\delta_{1})W_{2}(\delta_{2})$, where $W_{i}(x_{i})$   can be found from
equation (\ref{eqn38}). 
From equation (\ref{eqn13}) it follows that 
when $x_{1}\approx x_{2} \approx \mu$ the spherical polar angle $\theta$
 is close to
  the critical latitude defined by  $\theta =\arccos(\mu).$  
Accordingly, we describe this region as the region near the critical latitude.
\subsubsection{An expression for the eigenfunction near the critical
latitude}
In order to obtain an expression for the eigenfunction that is valid near the 
critical latitude we proceed as follows.
At first  we consider a region of the planet 
sufficiently far from rotational axis, where $\lambda (1-x_{1})\gg 1$.  
We start from the form of solution given by equation(\ref{eqn32})
in the form
\begin{equation}
W\propto (\pm(\mu^2-x_{2,1}^{2})^{-n/2}(1-x_{2,1}^{2})^{-1/4}\cos (\lambda
y_{2,1} +\phi_{2,1})\bar W(\delta_{1,2}),
\label{eqn320}
\end{equation}
where here and below
 the first index and upper sign (second index and lower sign)
correspond to the $(+)$ branch ($(-)$
branch).
The function $\bar W(y) $ satisfies equation
(\ref{eqn33}) with $m_{*}$ set to zero. As we discussed above the term
proportional to $m_{*}$ gives a correction which will be calculated
below. The fact that the function $\bar W(y) $ is normalised in order
to have the appropriate limit in the case of large $y$ is stressed by
the overbar.
We have
\begin{equation}
\bar W(y) =\sqrt {\pi\kappa \over 2}y^{(1-n)/2}J_{(n-1)/2}(\kappa y),
\label{eqn60}
\end{equation}
where  we recall that 
 $\kappa =\lambda/\sqrt{1-\mu^{2}}.$

Equation (\ref{eqn320}) is not valid near the critical latitude 
where  both $x_1$ and $x_2$ are close to $\mu.$
To obtain a modified form that is valid, the function
$\cos(\lambda
y_{2,1} +\phi_{2,1})$ must be replaced by a function that matches this
when $\lambda |y_{2,1}-\arccos(\mu)|$ is large but which takes  has the 
correct form  to result in the 
proportionality  of the eigenfunction,  as indicated above 
 to $\bar W(\delta_1)\bar W(\delta_2),$ when this is small.
It  can be seen that an 
expression for $W$ having the required properties can be written in the form
\begin{equation}
W_{\pm}=D_{*}^{-1/2}(1-x^{2}_{2,1})^{-1/4}(\pm(\mu^{2}-x_{2,1}^{2}))^{-n/2}
\bar W(\delta_{1,2})\Delta_{2,1}^{n/2}\bar W(\Delta_{2,1}), 
\label{eqn58}
\hspace {3mm} {\rm where}
\end{equation}
\begin{equation}  
\Delta_{2,1}=\pm\sqrt{1-\mu^{2}}(y_{2,1}-\arccos (\mu) ).
\label{eqn59}
\end{equation}
Here the quantity $D_{*}=D((1-\mu^{2})\mu)^{-n}$, where $D$ is defined in
equation (\ref{eqn21}) \footnote{Let us stress that the   $\Delta_{2,1}$
is not necessarily small contrary
to $\delta_{2,1}$ defined through  equation (\ref{eqn30a}).}. 
We remark that in the limit $\kappa
\delta_{1,2}\rightarrow \infty $ together with the limit
$\kappa
\Delta_{1,2}\rightarrow \infty $
we can use the 
asymptotic expansion of Bessel functions
\begin{equation}
J_{(n-1)/2}(\kappa y)=\sqrt{{2\over \pi \kappa y}}[\cos{(\kappa y -{\pi n\over 4})}
(\sum_{k=0}^{\infty} {A_{k}\over (\kappa y)^{2k}}) - \sin{(\kappa y -{\pi n\over 4})}
(\sum_{k=0}^{\infty} {B_{k}\over (\kappa y)^{(2k+1)}})],
\label{eqn61aa}
\end{equation}
where
\begin{equation}
A_{k}={(-1)^{k}\over 2^{2k} (2k)!}{\Gamma ({n\over 2}+2k)\over \Gamma ({n\over 2}-2k)},
\quad 
B_{k}={(-1)^{k}\over 2^{(2k+1)} (2k+1)!}{\Gamma ({n\over 2}+2k+1)\over \Gamma ({n\over 2}-2k-1)},
\label{eqn61bb}
\end{equation}
together with equation(\ref{eqn44}) 
to show that expression
(\ref{eqn58}) attains the required form specified in equations
$(\ref{eqn24})$ and $(\ref{eqn30})$.

\subsubsection{An expression for the eigenfunction near the pole and
 surface}\label{expr}
In the region close
to the pole of the planet we have $x_{2}\sim \mu$
and $\delta=1-x_{1} \ll 1.$  The discussion given above excluded
consideration   of this domain which needs to be considered separately.  
Close to the pole but away from the surface the solution is given by
equation (\ref{eqn25}).  From very similar considerations to those above,
close to the surface  where $|\lambda\delta_2|$ is of order unity or less
and to the  pole  where $1-x_{1} \ll 1,$ the solution is proportional, to within, in the limit of large $\lambda,$
a slowly varying proportionality factor,  to the product of $\bar W{(\delta_2)}$ and the
solution given by equation (\ref{eqn27})  which is valid near the rotation axis.
 Thus in this domain    $W\propto W_{a}(\delta )\bar W{(\delta_2)}.$ An 
an approximate solution for  $W_{pole}$ 
that  satisfies the required matching conditions and also reduces to the form (\ref{eqn58}) in the
limit $\lambda (1-x_{1})\gg 1$ can be written as 
\begin{equation}W_{pole}=(-1)^{k_{1}}D_{*}^{-1/2}(1-x_{1}^{2})^{-1/4}(x_{1}^{2}-\mu^{2})^{-n/2}
\sqrt{{\pi \lambda y_{1}\over 2}}J_{|m|}(\lambda y_{1})\bar W(\delta_{2}),\label{eqn61n}\end{equation}
where the factor $(-1)^{k_{1}}$ takes into account that when $k_{1}$ is odd $W_{a}$ and $W_{WKBJ}$ differ
by sign in the matching region, see equation (\ref{eqn44}).
  
\subsection{An approximate expression valid over the whole surface  domain}\label{dom}
We now use an interpolation procedure to combine  expressions derived above, that are valid in separate
domains inside the planet,  to form  single expressions that can be used over the whole domain.
To do this we  introduce a function $\eta [x] $ that 
is defined for   $x\in [0,1]$ and belongs to $C^{(\infty)}.$ 
We stipulate that  $\eta [x] $ decreases 
monotonically with $x$  and is such that  $\eta [x] \equiv 1$ when $x\in [0,1-x_{*}]$ 
and $\eta [x] \equiv 0$ when $x\in [x_{*},1]$
The quantity $x_*$  is a parameter such that $1/2 < x_{*} < 1.$ 
An explicit form of $\eta[x]$ 
as well as the value of $x_{*}$ are not important for our purposes.
This is because all representations in the different domains attain
a matching  asymptotic  form  in the planet interior and elsewhere.
  
 \subsubsection{An expression for $W_{-}$}
  
We may now  write down an approximate solution valid close to the surface
for all  $x_1\in [\mu,1]$
for  $W_{-}$ which we denote as $W_{-}^*$  as
\begin{equation}W_{-}^{*}=\eta[z_{1}]W_{-}+
(1-\eta[z_{1}])W_{pole},
\label{eqn62n}\end{equation}
where $z_{1}=(x_{1}-\mu)/(1-\mu)$ 
and we recall that $W_{-}$ in the above is to be obtained 
from equation (\ref{eqn58}).
The expression (\ref{eqn62n}) can be rewritten in another useful form, which explicitly shows that
the solution is separable close to the surface:
\begin{equation}
W_{-}^{*}=D_{*}^{-1/2}(1-x_{1}^{2})^{-1/4}(x_{1}^{2}-\mu^{2})^{-n/2}\tilde W_{1}(x_{1})\bar W(\delta_{2}),
\label{eqn62na}
\end{equation}
where
\begin{equation}
\tilde W_{1}=\eta [z_{1}] \Delta_{1}^{n/2}\bar W (\Delta_{1})+(1-\eta [z_{1}])
\sqrt{\pi \lambda y_{1}\over 2}J_{|m|}(\lambda y_{1}).
\label{eqn62nb}
\end{equation}

\subsubsection{An expression for $W_{+}$}

In this case we formulate an expression valid
close to the surface and for all $x_2 \in (0,\mu).$
Close to the equatorial plane and away from the critical latitude,  and $\kappa(\mu-x_2) =\kappa \Delta_{2} \gg 1$
 it is convenient to represent the 
function $W_{+}$ in terms of an asymptotic series in  ascending  powers of $(\kappa \Delta_{2})^{-1} $.  
Substituting the series (\ref{eqn61aa}) in (\ref{eqn58}) for $W_{+}$  we obtain:
\begin{equation}W_{eq}=D_{*}^{-1/2}(1-x_{2}^{2})^{-1/4}(\mu^{2}-x_{2}^{2})^{-n/2}
\tilde W_{eq}(x_{2})\bar W(\delta_{1}),
\label{eqn62nc}
\end{equation} 
where
\begin{equation}
\tilde W_{eq}(x_{2})=(-1)^{k_{2}}\cos \lambda (y_{2}-{\pi\over 2})
\sum_{k=0}^{\infty} {A_{k}\over (\kappa \Delta_{2})^{2k}} +(-1)^{(k_{2}+1)}
\sin \lambda (y_{2}-  \pi/ 2)
\sum_{k=0}^{\infty} {B_{k}\over (\kappa \Delta_{2})^{(2k+1)}}, 
\label{eqn62nd}
\end{equation}
where the coefficients $A_{k}$ and $B_{k}$ are given in equation
(\ref{eqn61bb}) and we take into account that 
$\kappa \Delta_{2}-{n\pi / 4}=\lambda (y_{2}-{\pi/ 2})
+\pi k_{2}$, according to equations (\ref{eqn44}) and (\ref{eqn59}). 

On the other hand
it is convenient to use the expression (\ref{eqn58}) directly  in the region close to the critical 
latitude. The expressions (\ref{eqn58}) and (\ref{eqn62nc}) can be combined with help of
the function $\eta [x]$  to provide a single function $W_{+}^{*}$, which is 
analogous to the function $W_{-}^{*}$ discussed above and can be used for $x_2 \in (0,\mu)$:
\begin{equation}
W_{+}^{*}=D_{*}^{-1/2}(1-x_{2}^{2})^{-1/4}(\mu^{2}-x_{2}^{2})^{-n/2}
\tilde W_{2}(x_{2})\bar W(\delta_{1}),
\label{eqn62nf}
\end{equation}
where
\begin{equation}
 W_{2}(x_{2})=\eta [z_{2}] \Delta_{2}^{n/2}\bar W(\Delta_{2}) + (1-\eta [z_{2}])
\tilde W_{eq}(x_{2}), 
\label{eqn62ng}
\end{equation}   
and $z_{2}=1-x_{2}/\mu$.

\subsection{Calculation of the mode norm}

In order to calculate different quantities related to a particular
mode we need an expression for the mode norm given by equation
(\ref{eq 16a}). One can show that when $\lambda $ is large enough the
interior of the planet  gives the dominant contribution  to the integrals 
determining the norm.  Thus the expression (\ref{eqn30}) for the eigenfunction 
is appropriate.  In addition 
one can simplify the expression (\ref{eq 16a}) by noting  that
eigenfunctions satisfy  equation (\ref{eq 4}) with the right hand side set to zero.  
Furthermore, the term involving the  operator $\bmth B$ in equation (\ref{eq 4}) 
may be neglected. This is  because it  does not contain second derivatives and 
 therefore  contributes 
a higher  order correction to the WKBJ approximation as  discussed above.
Additionally,  for the same reason,  only the terms
proportional to the second derivatives 
in the operators $\bmth A$ and $\bmth C$
 need to  be retained since these
terms give the leading contributions to the norm being 
proportional to $\lambda^{2}$.  Thus we have
\begin{equation}  
\sigma^{2}{\bmth A}W_{WKBJ}\approx {\bmth C}W_{WKBJ},
\label{eqn48} 
\end{equation}
and 
\begin{equation}  
N\approx 2(W_{WKBJ}|{\bmth
C}W_{WKBJ})=8\Omega^{2}\int _Vd\varpi dz\left ({\partial \over \partial z}F\right)^{2},
\label{eqn49} 
\end{equation}
where  in order to  evaluate  the norm (\ref{eq 16a}), 
we use the explicit form of $\bmth C$ given by equation (\ref{eq
6}) and,  after an integration by parts,  adopt equation  (\ref{eqn30}) for the eigenfunction.
 The quantity $F$ can be expressed in
the form
\begin{equation}  
F={1\over 2}(\cos (\lambda y_{+} -\Psi_{+})+\cos(\lambda y_{-}-\Psi_{-})), 
\label{eqn50} 
\end{equation}
where
\begin{equation} 
y_{\pm}=y_{2}\pm y_{1}, \quad \Psi_{\pm}={\pi \over 2}|m|+{\pi \over
  4}\pm {\pi \over 2}\lambda.
\label{eqn51} 
\end{equation}
Note that $y_{\pm}$ can be readily expressed in terms of coordinates
$(x_{1},x_{2})$ and $(\varpi,z)$:
\begin{equation}  
\cos y_{\pm} =x_{1}x_{2}\mp \sqrt{1-x_{1}^{2}}\sqrt{1-x_{2}^{2}}
=\mu z\mp (\sqrt{1-\mu^{2}})\varpi,
\label{eqn52} 
\end{equation} 
where we use equations (\ref{eqn13}). From equation (\ref{eqn52}) it
follows that the quantities $y_{\pm}$ are constant on characteristics
of equation (\ref{eq 4}), see equation (\ref{eqn23}). 
By differentiating (\ref{eqn50}) we then obtain
\begin{equation}
{\partial F \over \partial z}=- {\mu \lambda\over 2}
\left({\sin (\lambda y_{+} -\Psi_{+})\over \sin y_{+}}
+{\sin (\lambda y_{-} -\Psi_{-})\over \sin y_{-}}\right).
\label{eqn54} 
\end{equation}

In the limit of large $\lambda $ the integral of square of the
derivative (\ref{eqn54}) in (\ref{eqn49}) can be approximately
calculated taking into account that only average values of 
${\sin^{2} (\lambda y_{\pm} -\Psi_{\pm})}\approx {1/2}$ give
a significant contribution to the integral. In this way we obtain
\begin{equation}
\int _V d\varpi dz \left({\partial  F\over \partial z}\right)^{2}={\mu^{2}\lambda^{2}\over 8}I,
\label{eqn55} 
\end{equation}
where
\begin{equation}
I\equiv \int _V rdr d\theta \lbrace 
{1\over \sin^{2}y_{+}}+{1\over \sin^{2}y_{-}}\rbrace=2\pi.
\label{eqn56} 
\end{equation}
Note that in order to
evaluate the integral (\ref{eqn56}) we use the fact that from (\ref{eqn52}) we have
$\sin^{2}y_{\pm}=1-r^{2}\cos^{2}(\theta \pm \arccos(\mu))$. The 
integral is elementary and  most easily done by noting that it is easily shown to be independent of $\mu,$
and  can  accordingly  be evaluated setting  $\mu=1.$ 
Substituting
(\ref{eqn56}) into (\ref{eqn49}) we obtain a very simple expression
for the norm:
\begin{equation}
N={2\pi (\mu \lambda \Omega )^{2}}.
\label{eqn57} 
\end{equation}

\subsection{Calculation of the frequency correction $\sigma_{1}^{m}$}\label{sec3.9}

As follows from equation (\ref{eqn46}) the eigenfrequencies of the
modes are degenerate  with respect to change of sign of $m$ in the
approximation we use. The correction to the eigenfrequencies
accounting for the dependence on sign of $m$, $\sigma_{1}^{m}$, 
is calculated above, see equation (\ref{eqnn2}). As follows from this
equation, $\sigma_{1}^{m}$ is determined by the integral
\begin{equation}   
I=-\int _V d\varpi dz {\partial \rho \over \partial \varpi} W_{k}^{2}.
\label{eqn62}
\end{equation}

The integral (\ref{eqn62}) has contributions from the
interior of the planet where the mode eigenfunction is given by 
(\ref{eqn30}) and  also close to the surface where equations (\ref{eqn58})
and (\ref{eqn61n}) apply. 
Contributions to the integral close to the surface arise from both the $(+)$ and $(-)$
branches. Accordingly, we have $I=I^{WKBJ}+I^{+}+I^{-}$. At first let
us evaluate the contribution from the inner region, $I^{WKBJ}$. In
order to do this  we substitute (\ref{eqn30}) into  (\ref{eqn62}), thus
obtaining
\begin{equation}  
I^{WKBJ}=-\int  dr d\theta F^{2}{d \over d r}\ln \rho. 
\label{eqn63}
\end{equation}
 Since the quantity $F$
is rapidly  oscillating  we  use the average
value of $F^{2}$, $<F^{2}>=1/4$ in (\ref{eqn63}).  Hence
\begin{equation}  
I^{WKBJ}=-{\pi\over 4}\int_0^{1-\epsilon}dr {d \over d r}\ln \rho ={\pi\over
  4}\ln {\rho_{c}\over \rho_{*}}, 
\label{eqn64}
\end{equation} 
where $\rho_{c}$ is the value of the central density of the planet and
$\rho_{*} \ll \rho_{c}$ is a value of density close to the surface of
the planet $(r=r_*=1-\epsilon)$  above which the contribution to $I$  has to be
determined by  a separate treatment of the surface region.
 With help of equation (\ref{eqn21})
equation (\ref{eqn63}) can be rewritten in the form
\begin{equation} 
I^{WKBJ}={\pi\over
  4}(\ln {\rho_{c}\over D} -n\ln x_{*}),  
\label{eqn65}
\end{equation} 
where $x_{*}=1-r_{*}= \epsilon$ is the dimensionless distance from the
surface corresponding to the density $\rho_{*}$: $\rho_{*}=\rho
(x_{*})$.

Now let us evaluate the contribution to  integral from  the region close to the
surface. For definiteness, let us consider the $(+)$ branch where
$\delta_{1}=x_{1}-\mu $ is assumed to be small. An analysis of
contribution of the region close to the critical latitude to the
integral (\ref{eqn62}) shows that this contribution is small and can be
neglected. Thus, we can use an expression of the form (\ref{eqn32}),
  choosing 
$i=1$ and  $j=2$ there,  that 
correctly matches (\ref{eqn30}) in the interior. We substitute this in  (\ref{eqn62})
and change the integration variables from cylindrical coordinates 
to  pseudo-spheroidal coordinates,  using the fact  that 
\begin{equation} 
\varpi d\varpi dz = {x_{1}^{2}-x_{2}^{2}\over
  (1-\mu^{2})\mu}dx_{1}dx_{2}\approx {\mu^{2}-x_{2}^{2}\over
  (1-\mu^{2})\mu}d\delta_{1}dx_{2},
\label{eqn66}
\end{equation}
(see eg. Wu 2005a). Also, we use the approximate density profile given
by equation (\ref{eqn21}) and express the radial variable $x$ there in
terms of $\delta_{1}$ and $x_{2}$ with help of (\ref{eqn31}). In this
way we obtain
\begin{equation} 
I^{+}=n\int dx_{2}d\delta_{1}
(1-x_{2}^{2})^{-1/2}F_{2}^{2}\delta_{1}^{n-1}\bar W^{2}(\delta_{1}),
\label{eqn67}
\end{equation}
where $F_2=\cos (\lambda
y_{2} +\phi_{2}).$
The integral (\ref{eqn67}) should be evaluated in a volume bounded by
the surfaces $x=0$ and $x=x_{*}$.

 First let us evaluate the integral
 $I_{n}=\int d\delta_{1}
\delta_{1}^{n-1}\bar W(\delta_{1})^{2}$. Using the explicit expression  (\ref{eqn60})
for $\bar W(y)$, we obtain
\begin{equation} 
I_{n}={\pi \over 2}\int^{y_{*}}_{0}dy J^{2}_{(n-1)/2}(y).  
\label{eqn68}
\end{equation}
Here we note that the integral converges only for the case $n >
0$. The integration variable  $y=\kappa \delta_{1}$ with $y_{*}=\kappa \delta_{*}.$ 
The condition  $y=y_{*}$ defines the 
 surface  $x=x_{*}.$  From
equation (\ref{eqn31}) we have
\begin{equation} 
\delta_{*}= {(1-\mu^{2})\mu x_{*}\over \mu^{2}-x_{2}^{2}}.
\label{eqn69}
\end{equation}

The integral (\ref{eqn68}) logarithmically diverges when
$y_{*}\rightarrow \infty $. When $y_{*}$ is sufficiently large it can
be represented in the form
\begin{equation} 
I_{n}={1\over 2}\ln (B_{n}y_{*}),
\label{eqn70}
\end{equation}    
where the constant $B_{n}$ can be calculated numerically for a general
value of n\footnote{The
factor ${1\over 2}$ is determined by the form of asymptotic
expression of $J_{\nu}(y)\approx \sqrt {{2\over \pi
y}}\cos(y-{\pi\over 2}\nu -{\pi \over 4})$. Substituting this to
(\ref{eqn68}), integrating the result between two values of  $y=y_{1},y_{2}$,
then assuming that $y_{2} \gg y_{1}$ we obtain this factor.} . 
Substituting (\ref{eqn70}) to (\ref{eqn67}), using the averaged value
of $F^{2}_{2}$, $<F_{2}^{2}>=1/2$, we obtain 
\begin{equation} 
I^{+}={n\over 2}\int^{\mu}_{0}dx_{2}(1-x_{2}^{2})^{-1/2}\ln
(B_{n}y_{*}),
\label{eqn71}
\end{equation}
where a multiplicative factor of two has been applied  in order to account for 
the contributions from both the
upper and lower hemispheres.
Using equation (\ref{eqn69}) we can bring (\ref{eqn71}) to the form
\begin{equation} 
I^{+}={n\over 2}(\ln (B_{n}\kappa (1-\mu^{2})\mu x_{*})
\int^{\mu}_{0}dx_{2}(1-x_{2}^{2})^{-1/2}-\Phi^{+}(\mu)),
\label{eqn72}
\end{equation}
where 
\begin{equation} 
\Phi^{+}(\mu)=\int_{0}^{\mu}dx
(1-x^{2})^{-1/2}\ln (\mu^{2}-x^{2}).
\label{eqn73}
\end{equation}

The same approach can be used to evaluate the integral corresponding
to the $(-)$ branch with the result: 
\begin{equation} 
I^{-}={n\over 2}(\ln (B_{n}\kappa (1-\mu^{2})\mu x_{*})
\int^{1}_{\mu}dx_{1}(1-x_{1}^{2})^{-1/2}-\Phi^{-}(\mu)),
\label{eqn74}
\end{equation}
where 
\begin{equation} 
\Phi^{-}(\mu)=\int_{\mu}^{1}dx
(1-x^{2})^{-1/2}\ln (x^{2}-\mu^{2}).
\label{eqn75}
\end{equation}
Thus, the integral corresponding the the region close to the surface,
$I^{s}=I^{+}+I^{-}$ can be evaluated as 
\begin{equation} 
I^{s}={n\over 2}({\pi \over 2}\ln (B_{n}\kappa (1-\mu^{2})\mu x_{*})
-\Phi_{tot}),
\label{eqn76}
\end{equation}
where
\begin{equation} 
\Phi_{tot}=\int^{1}_{0}dx(1-x^{2})^{-1/2}\ln |x^{2}-\mu^{2}|.
\label{eqn77}
\end{equation}
It can be shown (eg. Prudnikov et al 1986) that the last integral does not
depend on $\mu$: $\Phi_{tot}=-\pi\ln 2$. Substituting this value to
(\ref{eqn76}), remembering that the total integral $I=I^{WKBJ}+I^{s}$ and
adding (\ref{eqn76}) to (\ref{eqn65}), we obtain the final expression
for the integral
\begin{equation} 
I={\pi \over 4}\ln \left[{\rho_{c}\over D}(4B_{n}\lambda \mu\sqrt{ 1-\mu^{2}})^{n}\right]. 
\label{eqn78}
\end{equation}
Note that the integral (\ref{eqn78}) does not depend on position
of the matching point $x_{*}$. Substituting it and the norm (\ref{eqn57}) to the expression for the
frequency correction (\ref{eqnn2}) we have
\begin{equation} 
\sigma_{1}^{m}=
{m\Omega \over \lambda^{2}}\ln \left[{\rho_{c}\over D}(4B_{n}\lambda \mu\sqrt{ 1-\mu^{2}})^{n}\right].
\label{eqn79}
\end{equation}
In Appendix B we show that the expression (\ref{eqn79}) has a correct
form in the limiting case of an incompressible fluid $n=0$.

The expression (\ref{eqn79}) is not valid when $|\mu| $ is
sufficiently close to $1$.  We recall that  one can prove that the absolute
value of any  eigenfrequency  must be less than or equal to $2\Omega $,( see
eg. PP). This condition may  be violated when
$\Delta \mu =1-\mu \ll 1$ and the correction (\ref{eqn79}) is added to 
the unperturbed  frequency (\ref{eqn46}). A similar constraint may be
obtained from consideration of the  assumptions leading to
(\ref{eqn70}). Indeed, the matching radius should obviously be  smaller
than the radius of the star, and, accordingly, $x_{*} < 1$. For small
values of $\Delta \mu $ this condition together with equation
(\ref{eqn69}) leads to 
\begin{equation}   
\delta_{*} < \mu,
\label{eqn80}
\end{equation}
where we assume that a 'typical' value of $x_{2}$ entering  (\ref{eqn69})
is of the order of $\sim \mu \approx 1$. On the other hand, for the
validity of the  asymptotic expression (\ref{eqn67}) we should have
$\kappa \delta_{*} > 1$, and therefore
\begin{equation} 
\delta_{*} < {\sqrt{\Delta \mu}\over \lambda}. 
\label{eqn81}
\end{equation} 
Combining inequalities (\ref{eqn80}) and (\ref{eqn81}) we obtain
\begin{equation} 
\Delta \mu > {C \over \lambda^{2}},
\label{eqn82}
\end{equation} 
where a coefficient $C > 1$ can be obtained from a more accurate
analysis. Thus, our simple approach is likely to be invalid for 
eigenfrequencies with absolute values sufficiently close to
$2\Omega$. Therefore, we discard unperturbed values (\ref{eqn46}) 
that lead to eigenfrequencies with absolute values larger than $2\Omega$ 
when the correction (\ref{eqn79}) is added.

The same analysis can be used to calculate 
the quantity $\sigma_{1}^{an}$ determining the correction to the
anelastic approximation. As follows from equation (\ref{eqn16d}) the
correction is proportional to an  integral very  similar in form  to that  given by
equation (\ref{eqn62}).  This integral also has contributions from the  surface and 
the interior where the standard  WKBJ solutions may be used.
These  contributions  are comparable and so they   should be matched at some radius.
In fact, it may be shown that the surface
contribution  
is equal to the surface contribution to  (\ref{eqn62}) given by 
equation (\ref{eqn76}). The internal WKBJ contribution is more complicated and should,
in general, be evaluated numerically. Therefore, for simplicity, we
adopt a different approach when dealing with the correction to the
anelastic approximation. We calculate  $\sigma_{1}^{an}$ numerically
using the expression (\ref{eqn16d}) for several 'global' eigenmodes (i.e. modes with 
a large scale distribution of perturbed quantities). Such modes are of especial importance in
applications of the formalism. For example, as discussed in PI, IP and in 
 PIN, those mainly determine the dynamic tidal response of the planet. Since 
$\sigma_{1}^{an}\propto \lambda^{-2}$, we expect that corrections to 
small scale WKBJ modes are less significant. They are, therefore, 
neglected. 

.

\section{ A comparison of analytical  and numerical results }\label{sec4}

\begin{figure}
\vspace{1cm}
\begin{center}
\includegraphics[width=9cm,angle=0]{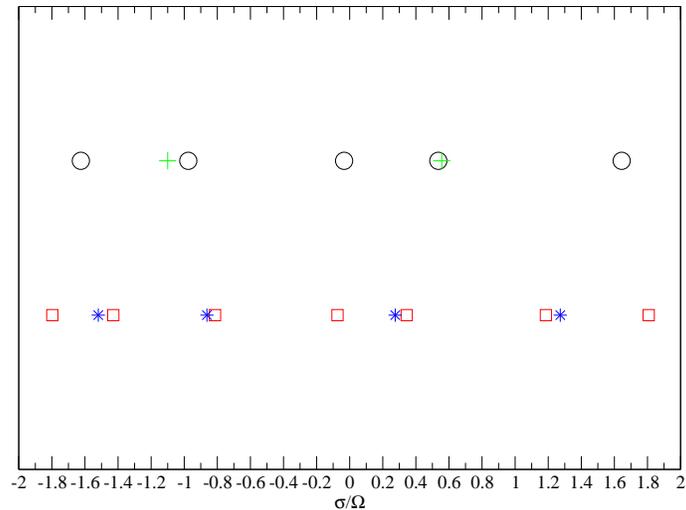}
\end{center}
\vspace{0.5cm}
\caption{Positions of the eigenfrequencies calculated in the WKBJ approximation
and the numerical results obtained by LF on the  $\sigma/\Omega $ axis
are plotted for the case $|m|=2$. The open circles and squares are
obtained from the WKBJ approach and the pluses and stars give
the results of the  numerical computations of LF.
The circles (pluses) correspond to
modes having $l=l_{min}$ while the open squares (stars)
correspond to modes with $l=l_{min}+1$.}
\label{f1}
\end{figure}

\vspace{1cm}

In this section we compare the frequencies  obtained from the approach
described above with those obtained by  a number of  authors
who have employed  a variety  numerical methods.

To obtain the values  from the above analysis we 
use  equation (\ref{eqn46}) for the eigenfrequencies, and, in the
case of non zero $m$,  add the expression for the correction to the
frequency given by (\ref{eqn79}).  We compare the results for polytropes  with  polytropic
indices $n=1$, and $1.5$. The quantities $B_{n}$
entering (\ref{eqn79}) for these cases were obtained numerically with help of equation
(\ref{eqn70}).  We found  $B_{1}\approx 14.4$ and $B_{3/2}\approx
5.97$. The range of allowed values of $k$ is found from
equation (\ref{eqn47}). Additionally, we shall discard modes, which have
$|\mu|$ close to $0$  or $1$ (see discussion below). We comment here that 
it follows from equation (\ref{eqn79})
that the frequency correction $\sigma^{m}_{1}$  becomes  undefined for such
values of $\mu.$

\begin{figure}
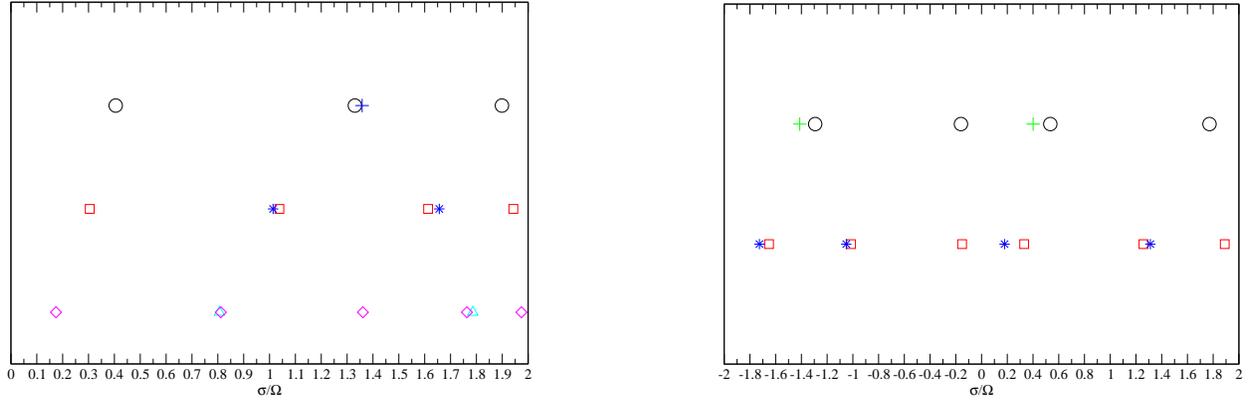

\vspace{1cm}
\begin{center}
\includegraphics[width=7cm,angle=0]{f2n.eps}
\hfill
\includegraphics[width=7cm,angle=0]{f3.eps}
\end{center}
\vspace{0.5cm}
\caption{As for Fig. \ref{f1} but the left panel illustrates  the
case $m=0$ and the right hand side panel corresponds to $|m|=1.$. 
In the left panel we also show the locations of the WKBJ modes
corresponding to $l=4$ (diamonds) and the locations of two eigenfrequencies
calculated by Dintrans $\&$ Ouyed
(2001)  (triangles).}
\label{f2}
\end{figure}

\vspace{2cm}

\begin{figure}
\vspace{1cm}
\begin{center}
\includegraphics[width=7cm,angle=0]{f4.eps}
\hfill
\includegraphics[width=7cm,angle=0]{f5.eps}
\end{center}
\vspace{0.5cm}
\caption{As for  Fig. \ref{f1} but for the case of
$|m|=3$ (left panel) and $|m|=4$ (right panel).}
\label{f3}
\end{figure}

\vspace{2cm}

\begin{figure}
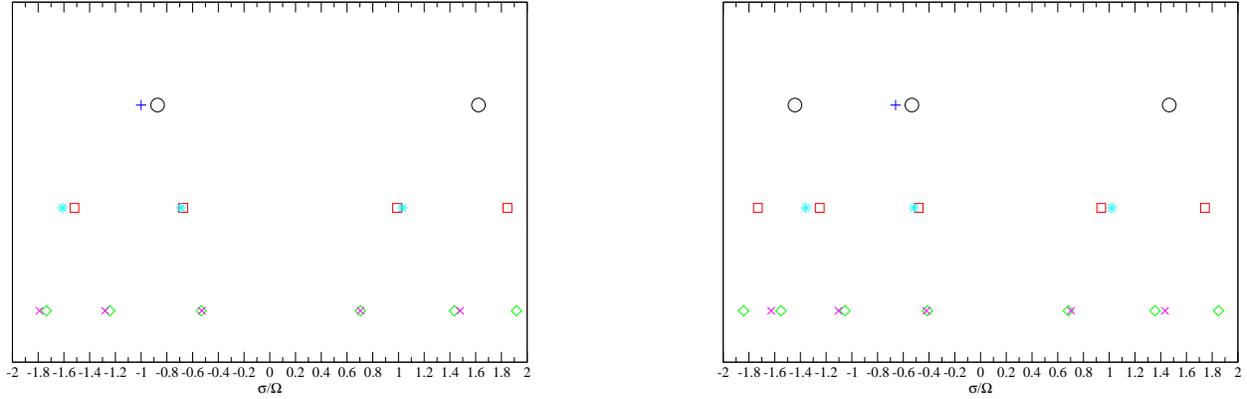

\vspace{1cm}
\begin{center}
\includegraphics[width=7cm,angle=0]{1LF.eps}
\hfill
\includegraphics[width=7cm,angle=0]{2LF.eps}
\end{center}
\vspace{0.5cm}
\caption{The result of comparison of the odd modes. Circles, 
squares and diamonds show positions of the WKBJ modes, corresponding 
to $l^{o}=0,1,2$, respectively. Pluses, stars and crosses give positions
of frequencies numerically calculated by LF.
The cases of
$|m|=1$ (left panel) and $|m|=2$ (right panel) are shown.}
\label{f1n}
\end{figure}

\vspace{2cm}

\begin{figure}
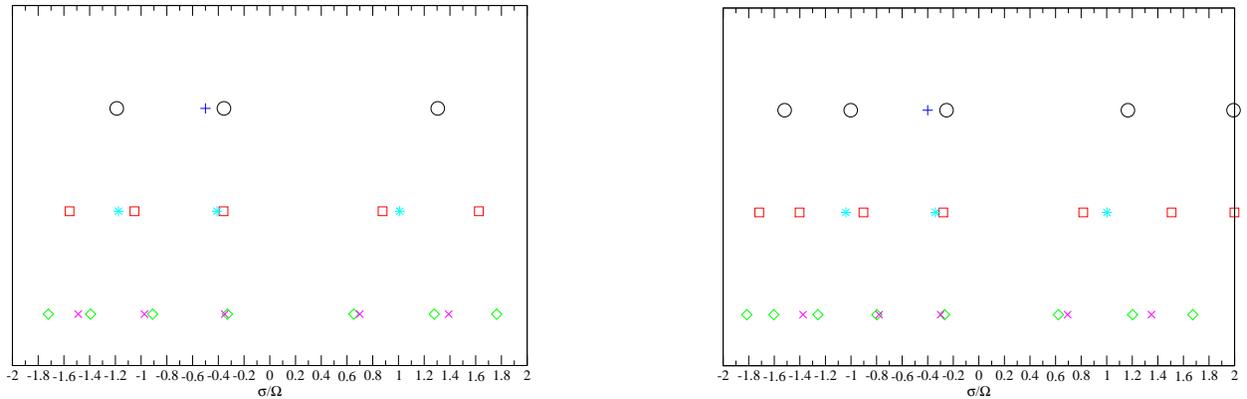

\vspace{1cm}
\begin{center}
\includegraphics[width=7cm,angle=0]{3LF.eps}
\hfill
\includegraphics[width=7cm,angle=0]{4LF.eps}
\end{center}
\vspace{0.5cm}
\caption{As for  Figs. \ref{f1n} but for the case of
$|m|=3$ (left panel) and $|m|=4$ (right panel).}
\label{f2n}
\end{figure}

\vspace{1cm}

Most  of the eigenvalues considered are for the model with $n=1$,
which has been  extensively discussed, (see eg. LF and  Dintrans $\&$ Ouyed,
2001). The amount of  attention paid to this model  is due to
the fact that it approximately reproduces the density distribution of
a cold coreless Jupiter mass  sufficiently below
the planet's surface. Some eigenvalues for  planetary models
(mostly the so-called 'global modes' with $m=2$ and $l=1$) were
calculated by IP.
The properties of the global modes in this case are very
similar to those of a polytrope with $n=1.$
 Since the different numerical
approaches essentially agree with each other for the
case $n=1$ with non zero values of $m,$
we compare our results with the numerical values obtained
by  LF.  

LF consider both even and odd modes and 
classify the mode order by an integer $l_{0}$, which is
related to integer $p$ defined in equation (\ref{eqn45n})
through  $l_{0}=p+|m|-1$. Let us remind that when the even modes are considered
$p=2l$, see equation (\ref{eqn45}), 
and for the odd modes $p$ is related to the integer $l^{o}$ classifying the odd modes
as $p=2l^{o}+1$, see equation (\ref{AA3}) of Appendix A.   
Using this definition, it is easy
to see that LF give eigenfrequencies of modes having  even symmetry
with $l=l_{min}=1$ and  $l=2$ for the case 
$|m| > 0$ and $l=l_{min}=2$ and $l=3$ for the case  $m=0.$
For the case of $m=1,$ we also considered the next order even modes with $l=4$
and compared eigenfrequencies with what  was obtained by  Dintrans $\&$ Ouyed
(2001). We also compare the WKBJ odd modes for $m=1,2,3$ and $4$ with the results 
of LS. The integer $l^{o}$, in this case, takes the values $l^{o}=l^{o}_{min}=0,1$ and $2$. 
For the case
$n=1.5$ the comparison of the WKBJ eigenfrequencies  is made with
the numerical results obtained by spectral methods in PI.

The results of  the comparisons for $n=1$ are shown in Figs. \ref{f1}- \ref{f2n}, where
positions of the eigenfrequencies within the allowed range $-2 < \sigma
/\Omega < 2$ are shown.  The WKBJ  
eigenfrequencies,
$\sigma $ are found from
\begin{equation}
\sigma =\sigma_{*}+\sigma_{1}^{m},
\label{add1}
\end{equation}
where $\sigma_{*} =2\mu $, and $\mu $, $\sigma^{m}_{1}$ are given by
equations (\ref{eqn46}), (\ref{eqn79}), respectively. In our
analytical investigation we assumed that the quantity $\mu$, and,
accordingly, $\sigma_{*}$, is positive, but allowed the sign of the azimuthal
number $m$ to be either positive or negative. For the purpose of this
section it is convenient to take  a different but equivalent point
of view and assume that the sign of $m$ is fixed: $m > 1$ and  allow
the quantities $\sigma_{*}$ to have either sign. The positive and negative signs
correspond to prograde and retrograde mode propagation with respect to
direction of rotation of the planet, respectively.

Let us first discuss the result of comparison of the even modes, which is presented in Figs. 
\ref{f1}-\ref{f3}.

In Fig. \ref{f1} illustrates
 the case with $m=2.$  This is  the most important value when one 
considers the problem of dynamical tides excited by a perturbing companion. 
 The WKBJ eigenfrequencies are shown as
open circles for $l=1$ and squares for $l=2$, respectively.  The
corresponding numerical values are indicated  by crosses and stars. As
seen in Fig. \ref{f1},  the numerical and analytical values show quite good
agreement with each other. 
This agreement is good even for the modes
with the smallest possible value of $l=1,$ which have a global
distribution of perturbed quantities over the volume of the star
and therefore might not be expected to be in any kind of  agreement with the results
of a WKBJ theory. This might  be explained by the rather large value of the
parameter $\lambda $ for these modes, being equal to $5.5.$ 
 From the discussion above, its inverse $\lambda^{-1}\approx 0.18$ is assumed to
be  a small parameter in our WKBJ expansions. There are, however,
three unidentified modes for the case $l=1$ as well as for $l=2$. These
modes have frequencies concentrating near the borders of the allowed
frequency range with $\sigma \approx \pm 2$ as well as in the region
close to $\sigma=0.$ It is possible that although these
correspond to global modes for the incompressible $(n=0)$ model,
these do not retain their character when $n$ is increased to $1$ and beyond
through a strong coupling to short wavelength modes nearby in eigenfrequency
(see discussion in section \ref{sec2.5}). 
 Discarding these modes, hereafter referred to as unidentified modes,
 the relative difference between the analytical
and numerical results is of order of or smaller than $10$ per cent.

In Fig. \ref{f2} we illustrate  the case with $m=0$ in the left panel
and the case with $m=1$ in the right panel.  For
$m=0$ the  eigenfrequency distribution is  symmetric with respect to
the reflection $\sigma \rightarrow -\sigma$, and, therefore, only positive
values of $\sigma $ are shown. As for the
previous case, when $m=0$ the agreement between the WKBJ values
classified as 'physical' and the results of the numerical study is
quite good, even when the 'global' mode with $\sigma \approx 1.3$
is considered,  the relative difference being of the order of $3$ per cent
for this mode . Again, this may be accounted for by the relatively
large value of $\lambda = 5.5$ for $m=0$ and
$l_{min}=2.$  Additionally,  on this plot we show the results of a calculation
by  Dintrans $\&$ Ouyed (2001). They calculated  five modes for
$m=0.$  Three of them have locations nearly the same as those
 obtained by LF. They are, therefore, not shown in the
plot.  As seen from the plot, the other two modes may be identified with
with the analytical modes having $l=4$. Note that the agreement gets
better with increasing   WKBJ  order $,l , $ as expected.

When $|m|=1$ the agreement is similar to the case with  $|m|=2$ apart from
the mode with $l=2$ and $k=2,$  where the WKBJ value $\sigma
\approx 0.33$ is approximately twice as  small  as the numerical
one. The reason for this disagreement is unclear to  us. Note,
however, that  there seems to be  a disagreement between numerical
methods in this case.  Only one of the modes with $l=2$ and
$|m|=1$ obtained by LF can be reliably identified with  an eigenfrequency 
 given by Dintrans $\&$ Ouyed (2001).
  However,  the 'global' modes with $|m|=1$ and
$l=1$ as well as all LF modes obtained for the case $m=0$ have
their counterparts in the results of Dintrans $\&$ Ouyed 2001.

In Fig. \ref{f3} we show the cases of relatively large values of
$|m|=3$ ( left  panel) and $|m|=4$ (right panel).
The agreement gets worse with increasing  $|m|$, and in the case of
the 'global' mode with $|m|=4$, $l=1$ and the numerical value 
$\sigma/\Omega \approx 0.6,$ the disagreement is of the  order of
$30$ per cent.  This may be explained by the fact that in our
theoretical scheme the value of $|m|$ is assumed to be much smaller
than the value of $\lambda$. However, for the mode with the largest
disagreement the ratio $|m|/\lambda \approx 0.53$ is not very small.

In Figs. \ref{f1n} and \ref{f2n} the comparison of the odd modes is made. 
The results are similar to the previous case. Apart from the presence of 
unidentified WKBJ modes, all numerically obtained frequencies have well 
identified WKBJ counterparts. The agreement is getting better with increase 
of the WKBJ order $l^{o}$ and is getting worse with increase of the azimuthal 
number $m$. Note a rather good agreement between the 'global' WKBJ and numerical 
modes corresponding to $l^{o}=0$. In fact, as was shown by PP (see also Papaloizou $\&$ Pringle 1977), 
eigenfrequencies and eigenfunctions 
of these modes can be calculated analytically giving very simple results
\begin{equation} 
\sigma=-{2m\Omega \over m+1}, \quad W=z\varpi^{m}.
\label{ne1}
\end{equation}
Note that neither eigenfrequencies nor eigenfunctions depend on the planet's density 
distribution in this case. 

In summary,  we point out that agreement between the numerical and
WKBJ frequencies  is unexpectedly good taking into account the fact
that the WKBJ theory should not,  strictly speaking, be applied to 
modes with such  small values of $\lambda .$  Apart from the existence of the
unidentified modes in the WKBJ scheme and the two 'physical' even modes and the global 
odd mode corresponding to $m=4$ with the  rather
large disagreements  alluded to above the agreement between analytical and
numerical results is of the order of or smaller than $15-20$ per cent for all
the remaining identified $27$ even and $35$ odd modes. As we shall see below,
good agreement is also found when  the spatial distribution 
of the modes is compared.

\subsection{Properties of eigenfunctions}

In order to calculate distributions of the quantity $W$ over the
volume of the planet we use equation (\ref{eqn30}) in the bulk of the
planet and equation (\ref{eqn58}-\ref{eqn62na}) close to the surface and smoothly
interpolate between the two regions. Since the function $\eta[x]$ defined in section \ref{dom} is
inconvenient for a numerical implementation we consider instead of it
a function, which is zero in the regions $[0, 1-x_{*}]$ and
$[x_{*},1]$ and represented as a ratio of two polynomials of $x$ in
the intermediate region, which are chosen in such a way to ensure
that several first derivatives are equal to zero in both points
$x=1-x_{*}$ and $x=x_{*}$.   

Let us stress that for self-consistency
 we use the frequency $\sigma_{*}$ (or $\mu$) 
as given by equation (\ref{eqn46})  in
those equations even when the frequency correction is non zero.
As above, the numerical results for  the  $n=1.5$
polytrope are taken from PI. The  WKBJ  results
for $n=1$ are compared with those obtained for a realistic model of
a planet  of one Jupiter mass. This model has  
a first order phase transition between  metallic
and molecular hydrogen, which has been  discussed, eg., in IP.

\vspace{1cm}

\begin{figure}
\begin{center}
\includegraphics[width=18cm,angle=0]{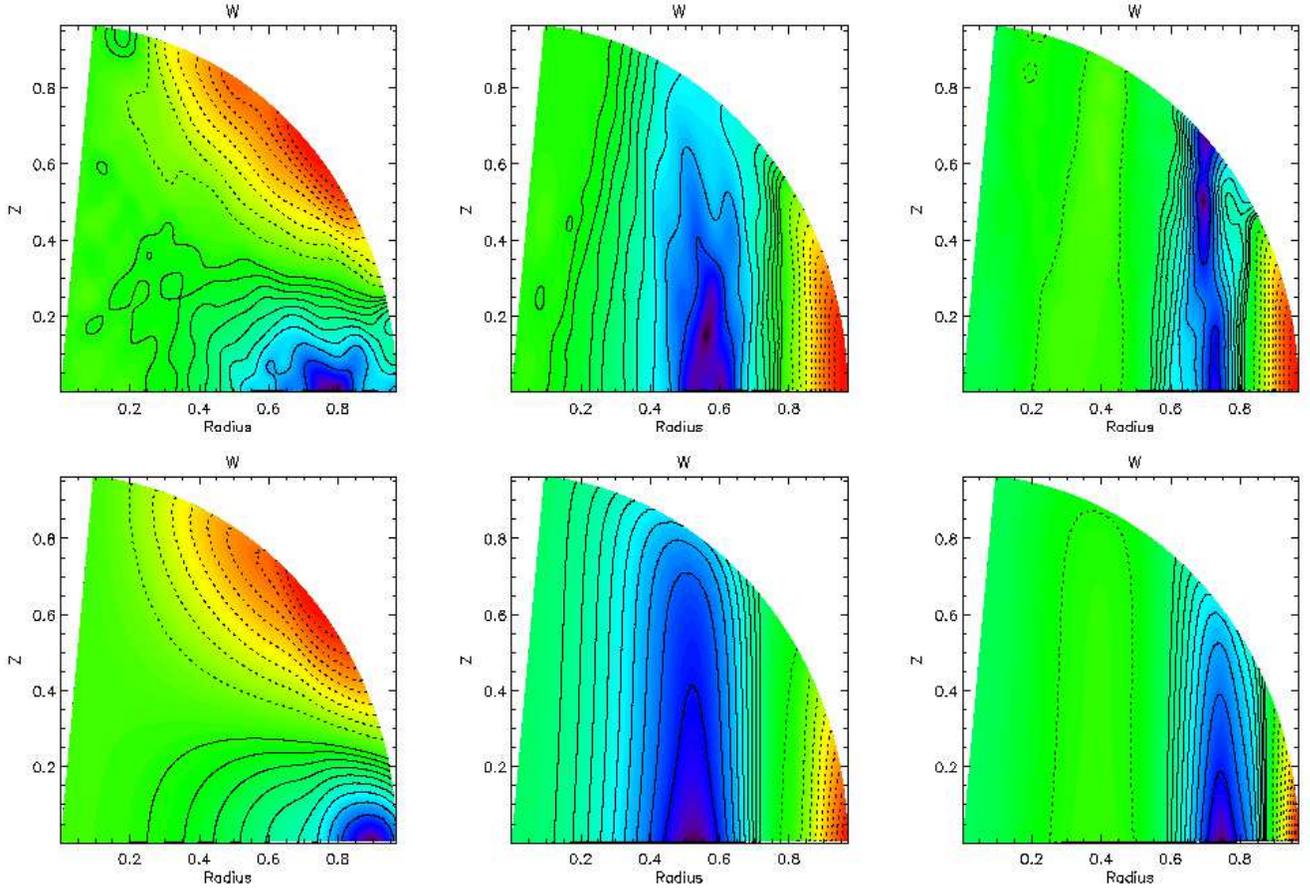}
\end{center}
\vspace{-5cm}
\caption{Distribution of $W$ over the surface of the star for the 
 $n=1.5$ polytrope. The upper plots are obtained by numerical
methods and the lower plots from the WKBJ theory. The
values of eigenfrequencies are given in the text.
From left to right the integers $l$ and $k$ determining the WKBJ
eigenfrequencies  are $l=1, k=0$, $l=1,k=1$ and
$l=2,k=2$ respectively.}
\label{ff1}
\end{figure}

\begin{figure}
\begin{center}
\includegraphics[width=18cm,angle=0]{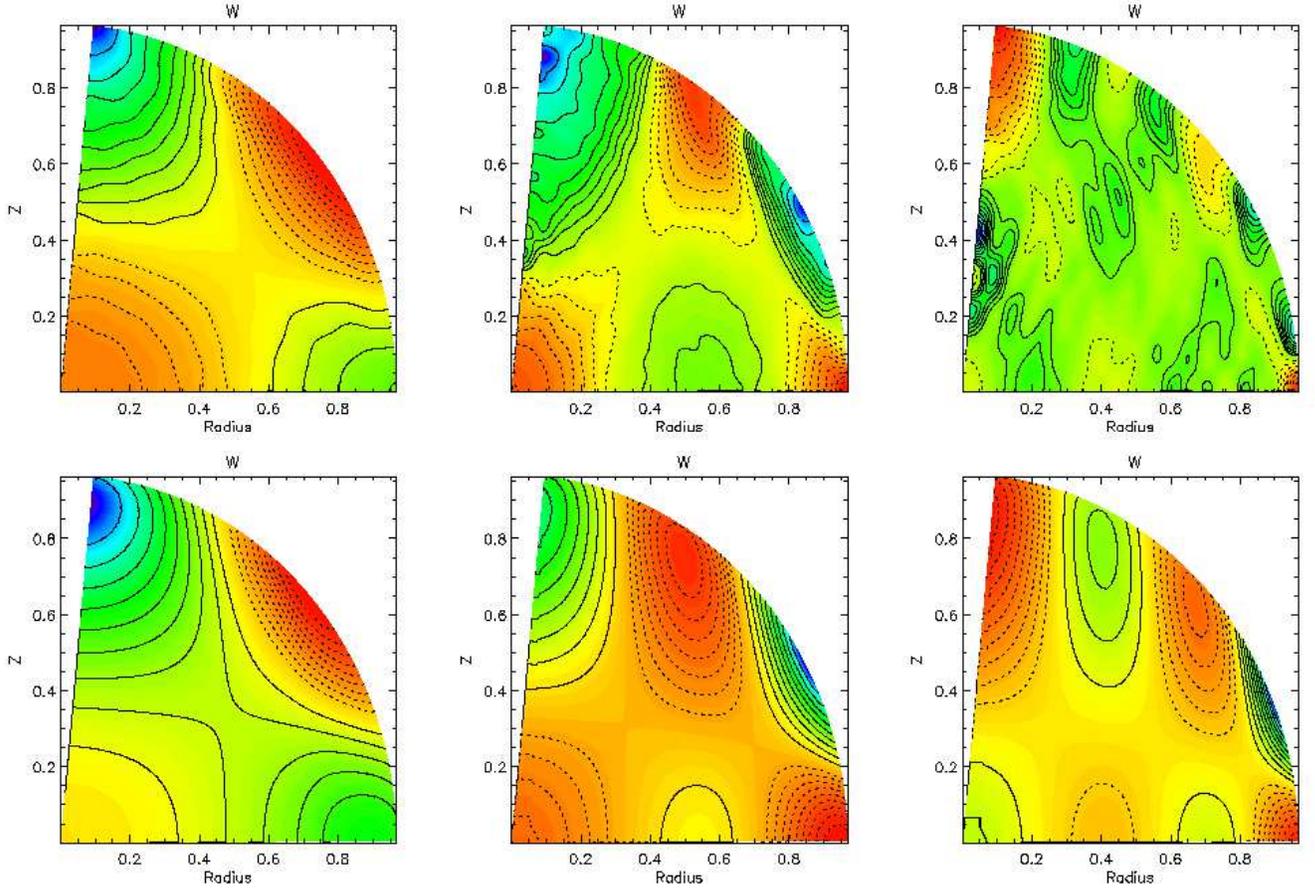}
\end{center}
\vspace{-5cm}
\caption{As for  Fig. \ref{ff1} but for $m=0.$ The upper plots
 represent numerical results  for a planet with a realistic
 equation of state, the lower plots are calculated from the WKBJ theory
 assuming that $n=1.$  The eigenfrequencies are given in the text.}
\label{ff2}
\end{figure}

\begin{figure}
\begin{center}
\includegraphics[width=18cm,angle=0]{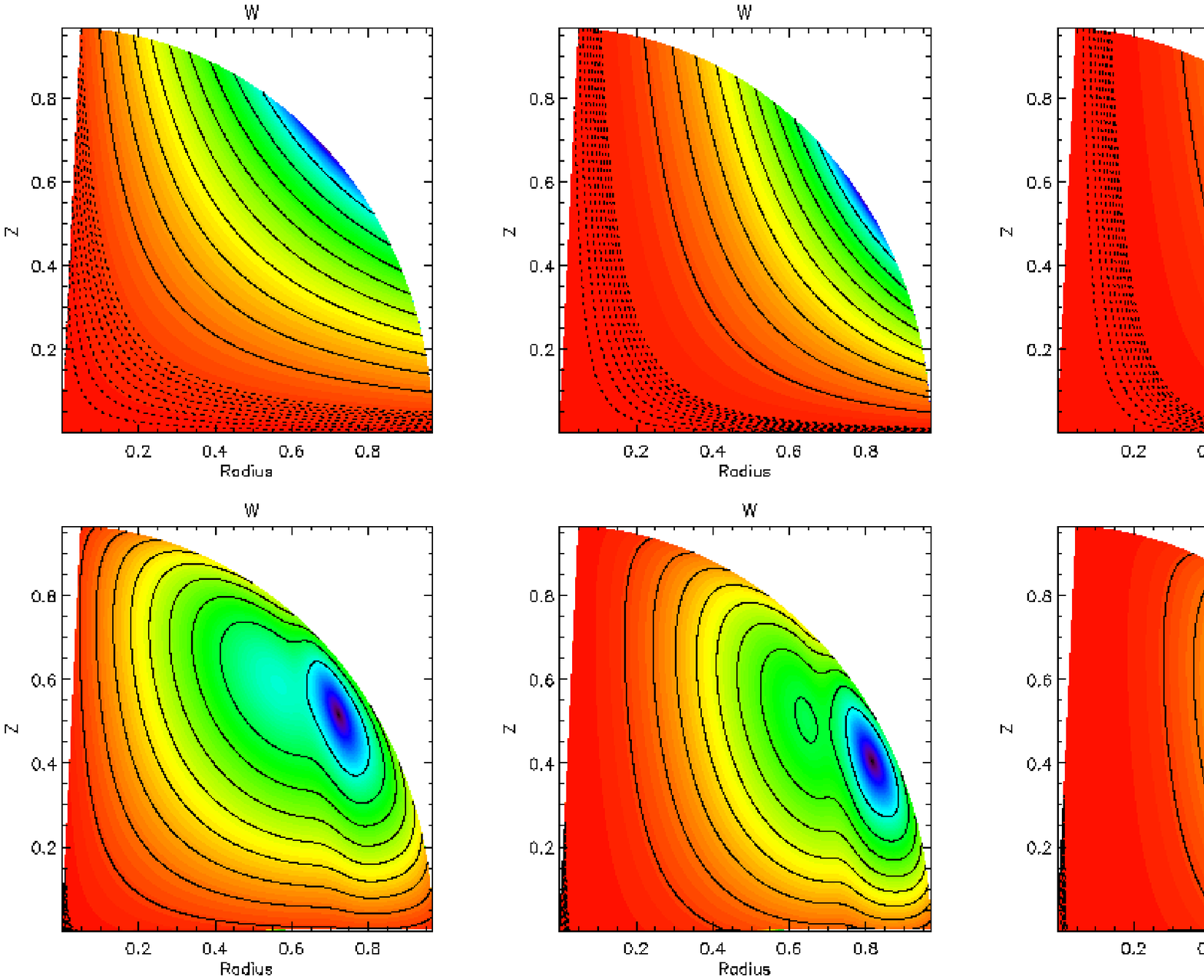}
\end{center}
\vspace{-5cm}
\caption{Distributions of $W$ for the global odd modes are shown. The azimuthal number 
 $m=1,2,3$ from left to right. The upper plots
 represent the analytical results given by equation (\ref{ne1}), the lower plots show the 
corresponding WKBJ counterparts.}
\label{ff3}
\end{figure}

A  comparison of the different results is  shown in Figs. \ref{ff1} and
\ref{ff2}. Note that in all cases shown in this section 
the same contour levels are used for the
numerical and analytical data. 

In Fig. \ref{ff1}   we show the  distribution of $W$ over the
planet's volume for the $m=2$ modes and $n=1.5$. Numerical results
are presented in the upper plots, which are taken from PI. 
These are for modes with $\sigma = -1.06$ ( upper left
plot), $\sigma = 0.67$ (upper middle plot) and $\sigma =0.67$ 
(upper right plot). The modes with $\sigma = -1.06$ and $\sigma = 0.67$
are the so-called two main global modes, according to 
PI.
They mainly determine transfer of energy
and angular momentum through dynamic tides induced by a parabolic encounter.
The respective WKBJ
counterparts have the smallest possible  WKBJ order $l=1$.
The corresponding analytic eigenfrequencies are $\sigma =-0.99$
($\sigma_{*}=-1.32$) and $\sigma=0.64$ ($\sigma_{*}=0.39$). The
distribution shown on the upper right plot may be identified with a
next order mode having $\sigma = 0.435$ and $\sigma_{*}=0.29.$  One can
see that there is a surprisingly good  agreement between the analytical and
numerical results. In particular, the retrograde mode represented on the left
hand side plots has a 'spot' in distribution at the angle $\approx
\pi/4$ with respect to the rotational axis. This agrees with position
of the critical latitude since $\arccos \mu =\arccos |\sigma_{*}/2|
\approx  0.27\pi $. The distributions shown on the middle and right
plots  correspond to prograde modes. They have a well pronounced
approximately vertical  isolines. The main global mode may be
distinguished from the mode corresponding to the next order by the  number
of nodes in the  horizontal direction, this being 
one in the case of the global mode and two for the next order mode.
We have checked that  similar  agreement  exists between
the WKBJ and numerical results corresponding to $n=1.$  Since the
distributions are quite similar they are not shown here.

For $m=0$ we compare the WKBJ results with calculations done by
a spectral method for a model of a planet of Jupiter size and
mass in Fig. \ref{ff2}. As in the previous case the upper plots correspond to
the numerical results. From left to right the numerical values of
the eigenfrequencies are $1.35$ (the main global mode), $1.01$ and
$0.79$. Their analytical counterparts have $l=2, k=1$, $\sigma =1.33$;
 $l=3, k=2$, $\sigma =1.03$ and  $l=4, k=3$, $\sigma =0.83$
respectively. Note that a more pronounced disagreement in eigenfrequencies 
corresponding to the mode represented on the right hand
side plot is mainly determined by the fact that this mode has
a   distribution concentrated near the surface of the planet,
where the  equation of state differs from that of a $n=1$ polytrope. One
can see that again there is very good  agreement between the
results. This is especially good for the main global mode represented
in the plots on the  left hand side. The agreement gets somewhat worse
moving from right  to left. This may be explained
by a number of factors such as inaccuracies of the numerical and
analytical methods as well as the physical effects determined by changes of the equation
of state in the  outer layers of the planet and the presence
of the phase transition. These factors mainly influence modes with a
small spatial structure while the large scale main global mode is
hardly affected by them.   

Finally we consider the global odd modes and 
compare the analytic distributions given by equation (\ref{ne1}) with the corresponding 
WKBJ distributions for $m=1,2$ and $3$ in Fig. \ref{ff3}. Although there is a disagreement in position 
of the spot close to the critical latitude, which is situated on the planet's surface in the case of
the exact analytic solutions and slightly interior to  the 
surface of the planet  in the case of the WKBJ distributions,  there is
a similarity in  the distributions in the  planet's interior. 
This is quite surprising since in this case the analytic distributions do not depend on the planet's 
structure at all while the WKBJ distributions are determined by the  density distribution close to the
planet's surface. 
  
\section{Discussion}\label{sec5}

\subsection{Overlap integrals}\label{sec5.1}

{ As we pointed out in the Introduction, integrals of the form
\begin{equation}
\hat Q_{k}={Q_{k}\over \sqrt N_{k}} \quad {\rm with}\quad 
Q_{k}= ({\rho \over c_{s}^{2}}W_{k}|\Phi),
\label{eqn84}
\end{equation} 
where $W_{k}$ corresponds to a particular eigenmode and $\Phi $ is 
some smooth function,
appear in astrophysical applications of the theory developed in this paper. 
In particular, as was discussed in PI and IP,
integrals of this type enter in expressions for the transfer of energy and
angular momentum transferred during  the periastron passage of a massive perturber. 
These apply to the case when the
spectrum of normal  modes is discrete
and they  involve integrals of form (\ref{eqn84}), where
$\Phi=P_{2}^{m}r^{2}$  with $P_{2}^{m}$   being the
associated Legendre function.
Assuming that  $W_{k}$ varies on a small spatial scale while the function $\Phi $ is smoothly varying 
such integrals may, in principal, be evaluated using our formalism  with
help of a theory of asymptotic evaluation of multidimensional integrals, see eg.  Fedoryuk (1987), Wong 
(1989). 

However, some important integrals of form (\ref{eqn84}) require an extension of our formalism, which
can provide a smooth matching of the solution close to the surface to the WKBJ solution in the inner part
of the planet that is valid  at  the next orders in inverse powers of $\lambda $. This is due to cancellations of leading 
terms in corresponding asymptotic series. Since this problem appears to be a rather generic one 
we would like to discuss it here in  more detail  for the important case 
when $\Phi =P^{2}_{2}r^2=3\varpi^{2}$. The overlap integral of this type determines excitation of the $m=2$ modes  which are the most
important for the tidal problem (eg. PI, IP and see also PIN). Explicitly, we have in this case   
\begin{equation}
Q_{k}= 3\int dV \left(\varpi^2 {\rho \over c_{s}^{2}}W_{k}\right),
\label{eqn84a}
\end{equation} 
where $dV=dz\varpi d\varpi $. Note that this integral must converge to zero in the incompressible limit 
$n\rightarrow 0$ as in this case it is well known
that inertial modes are not excited in the anelastic approximation. 
This fact, however, is not obvious for the integral written in the form 
(\ref{eqn84a}) because  close to the surface we have  
\begin{equation}
{\rho \over c_{s}^{2}}\approx nCx^{(n-1)},   
\label{eqn85}
\end{equation} 
with the constant $C$ converging to a nonzero value  as $n \rightarrow 0.$  Therefore,  
as the eigenfunctions are regular, the integral $Q_{k}/n$ has a logarithmic 
divergence at the surface of the planet as $n \rightarrow 0.$  This raises the possibility that the 
overlap integral might converge to a nonzero value or behave pathologically as the
incompressible limit is approached. 

In order to show that, in fact, this is not so and $Q_{k}(n\rightarrow 0)\rightarrow 0$ in a smooth
manner, let us consider 
some fiducial models having the property that the quantity  
\begin{equation} \omega_0^2 = -\frac{c^2_s}{r \rho }\frac{d\rho }{dr}\label{Good1} 
\end{equation}
is constant.  For models
in  hydrostatic equilibrium  under their own gravity, constancy of $\omega_0^2$ implies 
that ratio $M(r)/r^{3}$, 
where $M(r)$ is the mass interior to the radius $r,$  is constant. 
The model must accordingly be  incompressible.   
 Goodman $\&$ Lackner (2009) obtained a wider class of models
 in hydrostatic equilibrium   under a fixed quadratic gravitational potential.
 Because the potential is fixed independently of the mass distribution
 and there are no constraints on the equation of state, such models may be
 constructed for an arbitrary density distribution.
 
  Now let use consider the  integral     
\begin{equation}
Q_{fud}= 3\int dV \left (\varpi^2 {\rho \over c_{s}^{2}}\omega_0^2W_{k}\right).
\label{eqn84b}
\end{equation}
For the fiducial models described above this is identical
to the overlap integral (\ref{eqn84a}) where 
we note that  we may adopt natural units  such that the constant $\omega_0^2, $ 
which should be identified with the surface value of $GM(r)/r^3$ is equal to unity
in that case.  
More generally the integrand in (\ref{eqn84b}) can be transformed using equation of hydrostatic equilibrium
(\ref{Good1}) such that  
\begin{equation}
Q_{fud}= -3\int dV \left ({\varpi^2\over r} {d\rho \over dr}W_{k}\right)=-3\int dV \left (\varpi {\partial \rho \over
 \partial \varpi}W_{k}\right).
\label{eqn84c}
\end{equation}

Now let us consider equation (\ref{eq 41}) 
for free normal modes in the anelastic approximation  by setting $\sigma =\sigma_k$ and
the right hand side of this
equation to zero. 
Then we multiply it by $\varpi^2$, set $m=2$, and integrate over $dV$. After 
removing derivatives of $W_k$ by integrating
by parts, assuming that the density vanishes at the surface boundary,
it is easy to see that it follows from (\ref{eq 41})
that $Q_{fud}=0$ in the anelastic approximation provided $\sigma_k \ne 2\Omega.$
This means, in particular,  
that $m=2$ inertial waves cannot be excited in the  Goodman $\&$ Lackner (2009) models 
in this approximation as was found by these authors  when compressibility was fully taken into account
(see  PIN for additional discussion).

Using the fact that $Q_{fud}=0$ we may rewrite (\ref{eqn84a}) 
for models under their own self-gravity quite generally,
adopting natural units,  as
\begin{equation}
Q_{k}=Q_{k}-Q_{fud}= 3\int dV \varpi^2 {\rho \over c_{s}^{2}}\left(1-\frac{M(r)}{r^3}\right)W_{k}.
\label{eqn84d}
\end{equation} 
Taking into account that the factor in the brackets is proportional to $x$ for small $x$, and, accordingly,
in this limit the integrand is proportional to $nx^n$ we see that now the logarithmic divergence of $Q_k/n$
disappears and, therefore, it is clear from the representation (\ref{eqn84d}) that the overlap integral indeed  smoothly
tends  to zero in the limit $n\rightarrow 0$.

The theory of asymptotic evaluation of integrals of the 
form (\ref{eqn84d}) tells that the values of such integrals
are determined either by inner stationary points, where gradient of the WKBJ phase vanishes or contributions close
to the surface or other parts of the integration domain, where the WKBJ approximation is not valid. 
From the expression of the $W_{k}$ in the WKBJ regime (\ref{eqn30}) it follows that there are no stationary
points in the inner region of the planet. Considering the regions close to the surface it appears to be reasonable 
to assume that the leading contribution is determined by the region close to the critical latitude, where
a 'hot spot' is observed in distributions of $W_k$, see the previous section. 
In this region the quantities $\delta_1=x_1-\mu$ and $\delta_2=\mu-x_2$ are small. We can use them as new integration 
variables in (\ref{eqn84d}) with help of (\ref{eqn66}), separate the contribution of this region to the integral
by  introduction of the functions $\eta[\delta_{1,2}]$ defined in section 3.7 in the integrand, 
and decompose the quantities in front of $W_k$ in powers of $\delta_{1,2}$ taking into account that $x\propto 
\delta_{1}\delta_2$ and $\varpi^{2}\approx (1-\mu^{2})$ in the leading order. 
Assuming that $W_{k}\approx \bar W(\delta_{1})  
\bar W(\delta_{2})$, where $\bar W(y)$ is given by equation (\ref{eqn60}), it is easy to see that the leading
contribution to (\ref{eqn84}) is given by a symmetric combination of two integrals involving Bessel functions
\begin{equation}
Q_{k}\propto I_{1}(\delta_{1})I_{2}(\delta_{2})+ I_{2}(\delta_{1})I_{1}(\delta_{2}),
\label{eqn84g}
\end{equation} 
where
\begin{equation}
I_{1}(y)=\int_{y=0}dy \eta[y]y^{n}J_{(n-1)/2}(\kappa y), 
\quad I_{2}(y)=\int_{y=0}dy \eta[y]y^{(n+1)}J_{(n-1)/2}(\kappa y),
\label{eqn84j}
\end{equation} 
where it is assumed that $\eta[y >y_{*}]=0$, $y_{*}$ lies within the range of integration and 
$1/\kappa \ll y_{*} \ll 1$.
As was shown by Larichev (1973) the integral $I_{1}(y)=0$ for any particular form of the function $\eta[y]$. Thus,
the leading order contribution to the overlap integral from the surface region close to the critical latitude 
is equal to zero. In principal, one can look for the next order terms. However, in this case our simple approach
to the problem seems to be inadequate since eg. the assumption that $W_{k}$ can be represented as a product of 
two functions separately depending on the coordinates may be broken at this level, etc.. A more accurate 
approach is left for a future work. We note, however, that this cancellation means that the overlap integral should decay rapidly with increasing $\lambda,$ possibly 
  being  inversely proportional to a large power of $\lambda.$ This may qualitatively 
explain why a small number of relatively large
scale modes are significantly excited by dynamic tides, see PI, IP and PIN.}

\subsection{Conclusions}\label{Conclu}

In this paper we  have developed a WKBJ approximation, together
with a formal first order perturbation approach 
 for calculating
the normal modes  of a uniformly rotating coreless
planet under the assumption of a spherically symmetric structure.
Matching of the general WKBJ form valid  in the interior
to separable solutions valid near the surface resulted in
expressions for eigenfunctions
that were valid at any location within the planet 
together with an  expression for the associated eigenfrequencies
given in section \ref{sec3.5}.  Corrections as a result of density gradient
terms neglected in the initial WKBJ approach were also obtained from formal first
order perturbation theory.

The corrected WKBJ eigenfrequencies 
obtained using the WKBJ eigenfunctions
were compared with results obtained numerically by several different authors
and found to be in good agreement,  away from the limits of the
inertial mode spectrum  where identifications
could be made,  even for modes with a global structure.
We also compared the spatial forms of the eigenfunctions with those 
obtained using the spectral method described in IP finding similar good 
agreement.

This is consistent with the idea that these global modes can be identified
and that first order perturbation theory works even though they are embedded
in a dense spectrum.

In further support of this, the formal first order
perturbation theory developed here is subsequently  used to estimate corrections
to the eigenfrequencies
as a consequence of  the anelastic approximation and is then compared
with simulation results for a polytropic model with $n=1$
in PIN. These different approaches
are found to be in  agreement  for small enough rotation
frequencies and also indicates  that, as implied by
the simplified discussion in section \ref{anelastic} of this paper,
that  corrections as a result of the anelastic approximation are never very significant
for the models adopted.

Our results show that the problem of finding eigenfrequencies and
eigenvalues of inertial modes allows for an approximate analytical
treatment, even in the case of modes having a large scale distribution of
perturbed quantities. 

Although we consider only the case of a polytropic planet, our
formalism can be applied to a much wider context. Indeed, the approach
developed here is mainly determined by the form  of the density
 close to planet's surface, where we assume that it is
proportional to a power of distance from the surface. Thus, we expect
that our main results remain unchanged for any density distribution,
which is approximately power-law close to the surface. In particular,
according to our results, all models of type having approximately the same behaviour
of the density distribution close to the surface should have approximately
the same eigenspectrum.   

The formalism developed here can be extended for an approximate analytic
evaluation of different quantities associated with inertial modes,
such as overlap integrals characterising interaction of inertial waves
with different physical fields, growth rates due to the CFS instability
and decay rates due to various viscous interactions and non-linear
mode-mode interaction. It may provide a basis for a perturbative
analytic analysis of more complicated models, such as realistic
models of star and planets flattened by rotation or models of
relativistic stars.  
   
As we discussed above, for a given value of WKBJ order, $l,$ 
some modes are identified with  modes obtained  numerically  while
others remain unidentified. Eigenfrequencies of the unidentified modes
are always either situated close to the boundaries of the frequency
range allowed for inertial modes, $\sigma =\pm 2\Omega $ or situated
close to the origin $\sigma=0$. We believe that our theory is not
applicable to these modes, and they  develop  a
small scale  contribution controlled by the closeness of the  positions of 
their eigenfrequencies to $\sigma=0$ and $\pm 2\Omega, $
and thus effectively move to higher order than allowed for. 
Accordingly we de not consider these modes when
comparing our results with results of direct numerical calculations of the 
excitation of inertial waves due to a tidal encounter reported in PIN.

\section*{Acknowledgements}

We are grateful to the referee, Jeremy Goodman, for his comments, which led to
improvement of the paper. 

PBI was supported in part by RFBR grant 08-02-00159-a, by the
governmental grant NSh-2469.2008.2 and by the Dynasty Foundation.

This paper was prepared to the press when both
P.B.I. and J. C. B. P.  took part in the Isaac Newton programme 'Dynamics of Discs
and Planets'. 


\vspace{-0.4cm}

\begin{appendix}

\section{The eigenfrequencies of the 'odd' WKBJ modes}

In order to find the eigenfrequencies of modes odd with respect to reflection $z\rightarrow -z$
the phase $\phi_{2}$ should be chosen in such a way that the corresponding eigenfunctions 
are equal to zero at the planet equatorial plane, and, accordingly, $W(z=0)=W(y_{2}={\pi\over2})=0$.
From this condition and equation (\ref{eqn24}) we obtain 
\begin{equation}
\phi_{2}=-\lambda{\pi\over 2}+{\pi\over 2}+\pi s,
\label{AA1}
\end{equation}
where $s$ is an integer. Analogously to the case of even modes the phase $\phi_{1}$ is determined by
equation (\ref{eqn29}) and the form of solution close to the surface is determined by equations (\ref{eqn42}) and
(\ref{eqn43}). From these equations and equation (\ref{AA1}) we get the compatibility conditions analogous 
to conditions (\ref{eqn44})
\begin{equation} 
{\pi \over 4}n+\pi k_{1}=\lambda \arccos
(\mu)-{\pi\over 2}|m|-{\pi \over 4} \quad {\rm and} \quad
{\pi \over 4}n+\pi k_{2}=\lambda \left({\pi \over 2}-\arccos
(\mu)\right)-{\pi \over 2}-\pi s,
\label{AA2} 
\end{equation}  
and, solving (\ref{AA2}) for $\lambda $ and $\mu $ we find that
\begin{equation}  
\lambda=2l^{o}+|m|+n+{3\over 2},
\label{AA3} 
\end{equation}
where $l^{o}=k_{1}+k_{2}+s$, and that the expression for $\mu $ is given by equation 
(\ref{eqn46}), where (\ref{AA3}) should be used. 
Comparing equations (\ref{eqn45}) and (\ref{AA3}) we see that both even and odd modes 
can be described by the same expression for $\lambda $ provided that
\begin{equation}  
\lambda=p+|m|+n+{1\over 2},
\label{AA4}
\end{equation} 
where $p=2l$ for the even modes and $p=2l^{o}+1$ for the odd ones, respectively.

\section{The oscillation spectrum of an incompressible
fluid contained in a rotating spherical container in the WKBJ limit}\label{A}
In this appendix we show that the eigenvalues obtained from our WKBJ
analysis agree with the corresponding values for an incompressible
fluid contained in a rigid spherical container in the WKBJ limit.

\subsection{Eigenvalues for the incompressible case  in the WKBJ limit}

It is well known that  the spectrum of  normal modes for  an incompressible
fluid in a rotating spherical container can be found analytically, ( see eg.  Greenspan
1968, p. 64). The eigenfrequencies are determined from the equation
\vspace{4mm}
\begin{equation}
-mP^{|m|}_{s}(\mu)=(1-\mu^{2}){d P^{|m|}_{s}(\mu)\over d\mu},
\label{A1}
\end{equation}
\vspace{4mm} 
where $P^{m}_{s}(\mu)$ is a  Legendre function, $s$ is  an
integer, we recall that $\mu=\sigma/(  2\Omega)$ and note that
although it is inconsequential for the use of ({\ref A1}),
 Greenspan's  definition of $m$ has the opposite  sign to that 
used in this paper. 

We determine the spectrum in the WKBJ limit
from  (\ref{A1}). In this limit $s$ is a large parameter.
 For  our purposes it is important to retain all 
terms  of order of $O({s^{-1}})$ and larger.
 We  use  the asymptotic form of 
  Legendre functions in the limit of large $s$  in the form
\vspace{4mm}
\begin{equation}
P^{m}_{s}(\cos \phi )\propto {1\over \sqrt {\sin \phi}}\left\lbrace
\cos \left(\left(s+{1\over 2}\right)\phi -{\pi \over 4}+m{\pi \over 2}\right)+\left(m^{2}-{1\over
  2}\right){\cos ((s+{3\over 2})\phi +{\pi \over 4}+m{\pi\over
2})\over 2(s+{3\over2})\sin \phi}
\right\rbrace.
\label{A2}
\end{equation} 
\vspace{4mm}
\noindent Here we have  used  the well known properties of gamma functions to transform
the asymptotic  expression given in Gradshteyn $\&$ Ryzhik (2000)  to the form
(\ref{A2}).

 We now  substitute (\ref{A2}) into (\ref{A1}) after setting $\mu
=\cos \phi $ in that equation.  We then discard all terms $< O({s^{-1}}),$  thus  obtaining
$$\hspace{-3.5cm} -m\cos \left (\left(s+{1\over 2}\right)\phi -{\phi \over 4}+|m|{\phi \over 2}\right )\approx
s\sin\phi \sin \left(\left(s+{1\over 2}\right)\phi -{\pi \over 4}+|m|{\pi\over 2}\right)$$
\begin{equation}
+{1\over 2}\cos \left(\left(s-{1\over 2}\right)\phi -{\pi\over 4}+|m|{\pi\over2}\right)
+{1\over 2}(m^{2}-{1\over 4}\sin ((s+{3\over 2})\phi +{\pi \over
4}+|m|{\pi \over 2}).
\label{A3}
\end{equation}  

The quantity multiplying $s$  in the first term on the right hand side of (\ref{A3})  should
be close to zero in order that  this first term  be of the same order as the other terms on the right
hand side.  This condition gives
\begin{equation}
\phi=\pi { j +{1/4}-{|m| / 2}+\delta\over s+{1/ 2}}, 
\label{A4}
\end{equation} 
where $j$ is an integer and $\delta $ is a small higher  order correction.  Setting it to zero
we have
\begin{equation}
\mu_{0}=\cos \left( \pi { j +{1/4}-{|m| / 2}\over s+{1/ 2}}\right).  
\label{A5}
\end{equation}
This expression agrees with equation (\ref{eqn46}) when we
consider the limit of incompressible fluid.  To do this we  set 
$n=0$ in (\ref{eqn45}) and (\ref{eqn46}),  assume that $j-|m|=k$ and set
\begin{equation} 
\lambda =s+{1\over 2}.
\label{A6}
\end{equation}
Comparing (\ref{A6}) with (\ref{eqn45}) we obtain 
\begin{equation} 
s=2l+|m|.
\label{A7}
\end{equation}
One can show that the requirement that $s-|m|$ is an even number 
determines eigenmodes for which $W$ has  even symmetry with respect to
the reflection $z\rightarrow -z$, see eg. Greenspan 1968, p. 65.
Substituting (\ref{A4}) in (\ref{A3}) we get an expression for the
correction $\delta $
\begin{equation} 
\delta =-{m+{1\over 2}({3\over 4}+m^{2})\mu_{0}\over \lambda \sqrt{1-\mu_{0}^{2}}}.
\label{A8}
\end{equation}
Now we substitute (\ref{A8}) into (\ref{A4}) to obtain
\begin{equation} 
\mu=\mu_{0}+{\mu_{0}\over 2\lambda^{2}}({3\over 4}+m^{2})+{m\over \lambda^{2}}.
\label{A9}
\end{equation}
The last term 
\begin{equation} 
\mu_{1}^{m}={m\over \lambda^{2}}
\label{A10}
\end{equation}
gives the leading order difference in eigenvalues belonging to eigenfunctions 
with values of $m$ of opposite sign. It is  shown
below that the expression (\ref{A10}) agrees with (\ref{eqn79})
provided that (\ref{eqn79}) is evaluated in the limit $n \rightarrow 0$.    

\subsection{The limit  $n\rightarrow 0$ of the expression (\ref{eqn79})
for the frequency correction $\sigma_{1}^{m}$}\label{B}

If $n$ is directly set to zero,  the integral (\ref{eqn68}) diverges at
$y=0$. In order to find a limiting expression for
(\ref{eqn79}) it is necessary to consider $n$  to be very small
and take the limit  $n \rightarrow 0$ in the final expression only after 
potentially divergent terms have been  cancelled. 
Thus we assume that $n > 0$  but very small and integrate (\ref{eqn68}) by parts
to obtain
\begin{equation}
I_{n}={\pi \over 2}\int^{y_{*}}_{0}dy J^{2}_{\nu}(y)={\pi \over
2}y_{*}J^{2}_{\nu}(y_{*})-\pi 
\int^{y_{*}}_{0}dyyJ_{\nu}{d\over dy}J_{\nu},
\label{A11}
\end{equation}
where $\nu =(n-1)/2$ and we remark that for $n > 0$
$(yJ^{2}_{\nu}(y)) \rightarrow 0$ when $y\rightarrow 0$. We use 
the well known relation 
\begin{equation} 
y{d\over dy}J_{\nu}(y)=\nu J_{\nu}-yJ_{\nu+1}, 
\label{A12}
\end{equation} 
to eliminate the derivative of the Bessel function
and  obtain
\begin{equation} 
nI_{n}=\pi[y_{*}J^{2}_{\nu}(y_{*})+\int_{0}^{y_{*}}dy yJ_{\nu}(y)J_{\nu+1}(y)].
\label{A13}
\end{equation} 
It is easy to see that the expression on the right hand side does not
diverge when $n\rightarrow 0$. However, the integral entering
the right hand side of (\ref{A13}) diverges at large values of $y_{*}$. 
In order to deal with this divergence
we use the asymptotic expression for the Bessel function, $J_{\nu}^{a}(y)$,
valid at large values of its argument in the form,
\begin{equation}
J_{\nu}(y)\approx J^{a}_{\nu}(y)=\sqrt{{2\over \pi y}}\left \lbrace \cos\left (y-\nu {\pi
  \over 2}-{\pi \over 4}\right)-{1\over 2y}{\Gamma (\nu + 3/2)\over \Gamma (\nu
  -{1\over 2})} \sin \left(y-\nu {\pi
  \over 2}-{\pi \over 4}\right)\right \rbrace.
\label{A14}
\end{equation}
From equation (\ref{A14}) it follows that the boundary term in
(\ref{A13}) can be evaluated as
\begin{equation}
y_{*}J^{2}_{\nu}(y_{*})={1\over 2\pi}\left(1+\cos\left(2y_* -n{\pi \over 2}\right)\right)+O(y_*^{-1}).
\label{A14n}
\end{equation} 
Here we remark that because $y^* \sim \lambda x_*,$ where $ x_*$ is the dimensionless distance to the
surface and $\lambda$ is large, $y^*$ may be large when $x_*$ is small corresponding to
being close to the surface. Thus use of the asymptotic expansions of Bessel functions
for large values of their arguments can be justified.
In addition,  setting
$\nu =(n-1)/ 2$ and $\nu+1=n+1/ 2,$ (\ref{A14}) gives
\begin{equation}
J^{a}_{{n-1\over 2}}(y)J^{a}_{{n+1\over 2}}(y)\sim{1\over \pi y}\left(\sin \left(2y
-n{\pi \over 2}\right)+{n\over 2y}\right),  
\label{A15}
\end{equation}  
and
\begin{equation}
I_{*}(y_*)\equiv \int ^{y_*}_{0}dyy J^{a}_{{n-1\over 2}}(y)J^{a}_{{n+1\over 2}}(y)\sim -{1\over 2\pi}
\cos \left(2y_* -n{\pi \over 2}\right) + {n\over 2\pi}\ln y_*.
\label{A16}
\end{equation}
  
Now,  restoring complete precision, we can represent the integral entering (\ref{A13}) as
\begin{equation}
\int_{0}^{y_{*}}dy yJ_{\nu}(y)J_{\nu+1}(y)=I_{*}(y_{*})+C_{n},
\label{A17}
\end{equation}
where the quantity $C_{n}$ does not contain any divergences and can
be evaluated in the limit $n\rightarrow 0$ and $y_{*}\rightarrow \infty$. 

Setting $n=0$ in (\ref{A16}) and using the exact expressions 
for  the Bessel functions given by
\begin{equation}
J_{-{1\over 2}}=\sqrt{{2\over \pi y}}\cos y, \quad J_{{1\over 2}}=\sqrt{{2\over \pi y}}\sin y,
\label{A18}
\end{equation}
we find
\begin{equation}
 C_{0}=\int^{y_{*}}_{0}dyyJ_{-{1\over 2}}(y)J_{{1\over
2}}(y)+{1\over 2\pi}\cos 2y_{*}= {1\over 2\pi}.   
\label{A19}
\end{equation}

Now we use  (\ref{A19}),  (\ref{A16})  and (\ref{A17}) 
together with equations (\ref{A14n}) and  (\ref{A13}) to  obtain in
the limit of small $n$
\begin{equation}
I_{n}\approx {1\over n}+{1\over 2}\ln y_{*}.
\label{A20}
\end{equation}
Using  this together with  (\ref{eqn68}) and (\ref{eqn70}) we deduce that
\begin{equation}
B_{n}=e^{2/n}.
\label{A21}
\end{equation}
Substituting (\ref{A21}) in (\ref{eqn79}) and taking the limit
$n\rightarrow 0$ we get
\begin{equation}
\sigma_{1}^{m}={2m\Omega \over \lambda^{2}},
\label{A22}
\end{equation}
where we use the fact  that $\rho_{c}/D=1$ as  $n \rightarrow 0$. Recalling that
$\mu ={\sigma /( 2\Omega)}$ we see that (\ref{A22}) is equivalent to
(\ref{A10}).

\end{appendix}

\bsp

\label{lastpage}

\end{document}